\documentclass[12pt]{article}

\RequirePackage[OT1]{fontenc}
\usepackage{natbib}
\RequirePackage[colorlinks,citecolor=blue,urlcolor=blue]{hyperref}
\usepackage[english]{babel}
\usepackage{amsthm}
\usepackage{amsmath}
\usepackage{amsfonts}
\usepackage{amssymb}
\usepackage{graphicx}
\usepackage{enumerate}
\usepackage{color}
\usepackage{array}
\usepackage{psfrag,epsf}
\usepackage{url} 
\usepackage[export]{adjustbox}
\usepackage{bm}
\usepackage{titling}

\DeclareMathOperator*{\argmax}{arg\,max}

\def\frac#1#2{{\textstyle{#1\over#2}}}

\DeclareSymbolFont{AMSb}{U}{msb}{m}{n}
\DeclareMathSymbol{\Natural}{\mathbin}{AMSb}{"4E}
\DeclareMathSymbol{\Integer}{\mathbin}{AMSb}{"5A}
\DeclareMathSymbol{\Real}{\mathbin}{AMSb}{"52}
\DeclareMathSymbol{\Rational}{\mathbin}{AMSb}{"51}
\DeclareMathSymbol{\Imaginary}{\mathbin}{AMSb}{"49}
\DeclareMathSymbol{\Complex}{\mathbin}{AMSb}{"43} 
\DeclareMathSymbol{\Disk}{\mathbin}{AMSb}{"44} 
\def\bi{\begin{itemize}}
\def\ei{\end{itemize}}
\def\bd{\begin{description}}
\def\ed{\end{description}}
\def\ben{\begin{enumerate}}
\def\een{\end{enumerate}}
\def\bv{\begin{verbatim}}
\def\ev{\end{verbatim}}

\def\calD{{\mathcal D}}

\def\bar#1{{\overline{#1}}}

\def\E{{\rm E}}

\def\2to{{\ {\buildrel 2\over \longrightarrow}\ }}

\def\I1ton{{$I_1,\ldots,I_n$}}
\def\X1ton{{$X_1,\ldots,X_n$}}
\def\Y1ton{{$Y_1,\ldots,Y_n$}}
\def\Z1ton{{$Z_1,\ldots,Z_n$}}
\def\R1ton{{$R_1,\ldots,R_n$}}
\def\e1ton{{$e_1,\ldots,e_n$}}
\def\t1ton{{$t_1,\ldots,t_n$}}
\def\x1ton{{$x_1,\ldots,x_n$}}
\def\y1ton{{$y_1,\ldots,y_n$}}
\def\z1ton{{$z_1,\ldots,z_n$}}

\newtheorem{defn}{Definition}

\newtheorem{prop}[defn]{Proposition}

\newcommand{\ffrac}[2]{\ensuremath{\frac{\displaystyle #1}{\displaystyle #2}}}

\begin{document}
\thispagestyle{empty}
\baselineskip=28pt
\vskip 5mm
\begin{center} {\Large{\textbf{Efficient Modeling of Spatial Extremes over Large Geographical Domains}}}
\end{center}



\baselineskip=12pt
\vskip 5mm

\begin{center}
\large
Arnab Hazra$^{1}$, Rapha{\"e}l Huser$^2$, and David Bolin$^2$
\end{center}

\footnotetext[1]{
\baselineskip=10pt Department of Mathematics and Statistics, Indian Institute of Technology Kanpur, Kanpur 208016, India. \\ E-mail: ahazra@iitk.ac.in}
\footnotetext[2]{
\baselineskip=10pt Computer, Electrical and Mathematical Sciences and Engineering (CEMSE) Division, King Abdullah University of Science and Technology (KAUST), Thuwal 23955-6900, Saudi Arabia.}

\baselineskip=17pt
\vskip 4mm
\centerline{\today}
\vskip 6mm

\begin{center}
{\large{\bf Abstract}}
\end{center}

Various natural phenomena exhibit spatial extremal dependence at short spatial distances. However, existing models proposed in the spatial extremes literature often assume that extremal dependence persists across the entire domain. This is a strong limitation when modeling extremes over large geographical domains, and yet it has been mostly overlooked in the literature. We here develop a more realistic Bayesian framework based on a novel Gaussian scale mixture model, with the Gaussian process component defined by a stochastic partial differential equation yielding a sparse precision matrix, and the random scale component modeled as a low-rank Pareto-tailed or Weibull-tailed spatial process determined by compactly-supported basis functions. We show that our proposed model is approximately tail-stationary and that it can capture a wide range of extremal dependence structures. Its inherently sparse structure allows fast Bayesian computations in high spatial dimensions based on a customized Markov chain Monte Carlo algorithm prioritizing calibration in the tail. We fit our model to analyze heavy monsoon rainfall data in Bangladesh. Our study shows that our model outperforms natural competitors and that it fits precipitation extremes well. We finally use the fitted model to draw inference on long-term return levels for marginal precipitation and spatial aggregates.

\baselineskip=16pt

\par\vfill\noindent
{\bf Keywords:} Censored inference; Extremal dependence; Heavy rainfall; Low-rank spatial process; Spatial extremes; Spatial-scale mixture; Stochastic partial differential equation.\\

\pagenumbering{arabic}
\baselineskip=24pt

\newpage

\section{Introduction}
\label{sec:intro}

Over the last decades, extreme weather and climate events, such as the 2003 or 2011 European heatwaves \citep{stott2004human, zhong2021modeling}, {extreme rainfall from Hurricane Harvey in Texas, US, in 2017 \citep{risser2017attributable, van2017attribution}}, the extreme winter storm 2022 which caused chaos and life-threatening hazards across the US, or the 2022 Pakistan floods that killed more than 1\,700 people and affected a third of the country, 
just to name a few examples, have raised major concerns among scientists and various stakeholders about the serious climate risks and challenges that our society faces more than ever on a global scale. Climate risk assessment is indispensable in many practical applications \citep{kharin2007changes, ghosh2012lack,zscheischler2020typology} and is especially relevant in the context of climate change, which exacerbates concerns about the (often aggravated) frequency, intensity, and spatial extents of climate extremes; see the International Panel on Climatic Change (IPCC) reports \citep{masson2021ipcc} and the 2021 COP26 conference goals \citep{calliari2020digital}. 
Recently, \cite{chen2018recent} have emphasized the need to build resilient statistical models for analyzing climate extremes. However, existing extreme-value models are often deliberately and conveniently applied to low-dimensional spatial datasets observed over relatively small regions \citep{smith1990max, padoan2010likelihood, davison2019spatial, huser2020advances}, mainly for computational reasons, but also because classical spatial extreme-value models often lack flexibility to capture key data characteristics realistically. 
{With the increasing availability of massive datasets (e.g., satellite-derived data products, climate model outputs, wide in-situ observation networks)}, it is often needed to model complex high-dimensional extremes observed over large geographical regions, requiring models that transcend the classical extreme-value paradigm.

To model high-resolution spatial extremes datasets over large domains, two general criteria need to be fulfilled: (i) first, key marginal and dependence features that determine and impact risk assessment over the domain must be captured accurately using realistic statistical models; (ii) and second, fast, scalable, and statistically efficient inference methods are required. In the context of modeling precipitation extremes, the first criterion requires flexible, yet theoretically well-supported models that capture heavy upper tails, short-range extremal dependence, mid-range extremal independence, and long-range full independence (as motivated by the first law of geography). Mathematically, extremal dependence between two random variables $Y_1 \sim F_1$ and $Y_2 \sim F_2$ may be described through the tail correlation coefficient $\chi = \lim_{u \rightarrow 1} \chi_u$, where
\begin{equation}\label{eq:chi}
    \chi_u =  {\rm Pr} \left\{Y_1 > F_1^{-1}(u) \mid Y_2 > F_2^{-1}(u)\right\};
\end{equation}
here, $\chi \in (0, 1]$ indicates asymptotic dependence (AD) and $\chi = 0$ indicates asymptotic independence (AI). For a spatial process, the $\chi$-measure between two locations $\bm{s}_1$ and $\bm{s}_2$ is defined by a function $\chi(\bm{s}_1, \bm{s}_2)$. For the first criterion to be fulfilled, we thus need parametric models that are ``expressive'' enough to capture a wide range of $\chi(\bm{s}_1, \bm{s}_2)$ values and sub-asymptotic behaviors (changing as a function of distance $\|\bm{s}_1-\bm{s}_2\|$) and that lead to joint tail characteristics that can be theoretically derived in closed form. For the second criterion to be fulfilled, we need models that are constructed with a \emph{sparse} probabilistic structure leading to simpler computations or likelihood functions without sacrificing the model's key properties. For aggregated spatial risk assessment over complex domains, a third criterion may also be desired: feasible and fast (un)conditional simulation from the model.

Classical spatial extreme-value models usually fall short in at least two, if not the three above criteria. They can essentially be divided into two broad categories. The first category comprises models for site-wise block-maxima using max-stable processes \citep[see, e.g.,][]{smith1990max,schlather2002models,kabluchko2009stationary,reich2012hierarchical,opitz2013extremal}. These max-stable models have several drawbacks \citep{huser2020advances,huser2024time}: {likelihood-based inference is not easily scalable to high dimensions \citep{castruccio2016high, Huser.etal:2024}}, and max-stable models may not be realistic over large domains, as they have a rigid dependence structure that exhibits either AD or exact independence, and they usually assume AD across the entire spatial domain. The second category comprises models for peaks over high thresholds that are often modeled using generalized Pareto processes \citep{dombry2015functional, thibaud2015efficient, de2018high}. There is a one-to-one correspondence between max-stable processes and Pareto processes, and both are related to each other in terms of a point process representation \citep[see, e.g.,][]{thibaud2015efficient,davison2019spatial}. In applications, though, these two approaches differ fundamentally in their definition of an ``extreme event'' (block maxima or threshold exceedances), which has practical implications. The main drawback of Pareto processes is, again, their limited applicability in high-dimensional problems, now due to multivariate censoring of low non-extreme values. Some progress has recently been made to fit high-dimensional multivariate Pareto distributions using less efficient gradient scoring rules \citep{de2018high}, using black-box likelihood-free neural Bayes estimators \citep{ Richards.etal:2023, Zammit-Mangion.etal:2024} or by exploiting conditional independence graphical constructions \citep{engelke2020graphical}. However, currently, there is no spatial version of the latter sparse graphical multivariate Pareto models. Moreover, Pareto processes cannot capture full independence (as their support is not a product space), and have limitations similar to max-stable processes in terms of their tail structure. In contrast to max-stable and Pareto processes, the conditional spatial extremes model of \cite{wadsworth2019higher}, motivated by asymptotic theory through conditioning upon the value at a particular site being extreme, is applicable in higher-dimensional problems. However, the main drawback of this approach is that there is no unifying ``unconditional representation'' in the sense that different conditional models, obtained by considering different conditioning sites, cannot be theoretically reconciled in the AI case. Thus, this model lacks a convenient, ``physical'' interpretation. Moreover, as originally defined, the model does not have a sparse probabilistic structure, but see the recent sparse generalizations by \citet{Vandeskog.etal:2022} and \citet{simpson2023high}, which trade some model flexibility against computational speed. 

The drawbacks of the above classical extreme-value and conditional models have motivated the development of ``sub-asymptotic'' models based on location and/or scale mixtures of Gaussian processes \citep{huser2017bridging, morris2017space, krupskii2016factor, hazra2018semiparametric, hazra2020multivariate}. The latter are often more flexible in their tail structure and computationally more pragmatic to use than classical approaches; see \cite{huser2020advances} for a review. These models (or extensions thereof) can, in some cases, conveniently capture AD or AI within the same modeling framework \citep[e.g.,][]{huser2017bridging,huser2019modeling}. However, they are very often constructed from common latent factors that affect all spatial locations simultaneously \citep[e.g.,][]{krupskii2016factor}, which induces long-range dependence and implies that complete independence cannot be captured, even at an infinite distance, a major limitation with large geographical domains. Related to this, these models usually assume either AD for all pairs of sites, or AI for all pairs of sites, but cannot capture a change in extremal dependence regime as a function of distance, which is often a more realistic assumption. One exception is the model proposed by \cite{morris2017space}. Following \cite{kim2005analyzing}, \cite{morris2017space} consider random spatial partitions of the spatial domain based on Voronoi tessellations \citep{green1978computing}, and assume independent random location and scale terms within each subdomain, which induces long-range extremal independence when the number of partitions is unconstrained. However, in practice, \cite{morris2017space} fix the latter in their algorithm, which defeats the theoretical benefit of this model. Moreover, Bayesian inference based on Markov chain Monte Carlo (MCMC) methods requires the partition indices to be imputed for each spatial location at each MCMC iteration, which considerably slows down the convergence of Markov chains.
Another model based on Cauchy kernel convolution processes with compactly-supported kernels has recently been proposed by \citet{Krupskii.Huser:2021} to capture short-range extremal dependence and long-range independence, but its likelihood function is intractable, which prevents Bayesian inference and requires less efficient moment-based estimation.

Our proposed approach differs from classical extreme-value models and solves the above modeling and computational challenges by building upon and extending Gaussian scale mixtures. The reason why we consider these models as a starting point is that they have a natural link with standard geostatistics, and they are easy to simulate from (both unconditionally and conditionally), while typically allowing for relatively fast inference. Moreover, scale mixture models have been shown to capture AI and AD depending on the tail behavior of the mixing variable \citep{huser2017bridging}. One especially interesting example, the so-called ``HOT model'' \citep{huser2017bridging}, is detailed in Section~\ref{methodology}. The HOT model is obtained by multiplying a standard Gaussian process with a random scale variable whose distribution bridges two distinct families, namely the Pareto and Weibull distributions, which lead to AD and AI, respectively. We here develop the spatial HOT model, or ``SHOT model'' in short, which improves the HOT model in several important ways: (i) First, we replace the single latent random scale with a spatial random process, in order to create local (rather than global) shocks. This construction realistically reflects the behavior of many natural phenomena, such as extreme rainfall events, which tend to be spatially localized. Unlike the HOT model, our proposed SHOT model with spatially-varying shocks can capture AD at short distances, AI at larger distances, and complete independence at infinite distances. (ii) Second, for computational reasons, we assume that the random scaling process is of low-rank structure, built from well-chosen basis functions placed on a mesh, with random coefficients. Although our construction is nonstationary and combines basis functions and random coefficients non-linearly, the core model remains approximately tail-stationary. (iii) Third, we take the underlying Gaussian process as a Mat\'ern-like Gaussian Markov random field \citep[GMRF,][]{rue2005gaussian}, obtained as the solution of a stochastic partial differential equation \citep{lindgren2011explicit} on a fine mesh, which yields a sparse probabilistic structure, so that computations are scalable to high dimensions. (iv) Fourth, while the model HOT was originally defined as a copula model, thus disregarding marginal distributions, {we here adopt a unified approach for the modeling and estimation of margins and spatial dependence, which facilitates accurate uncertainty quantification in a fully Bayesian framework.} We also incorporate spatially-varying location parameters for additional flexibility in fitting nonstationary marginal distributions. Bayesian inference for our proposed model can be performed using MCMC methods, and we further show how the inclusion of a nugget effect in the construction of the underlying GMRF yields fast imputation of censored observations below a high threshold, which allows us to fit the model to multivariate threshold exceedances efficiently. {The usefulness of the nugget effect in hierarchical Bayesian modeling was demonstrated in the context of censored geospatial data 
 \citep{banerjee2003hierarchical, sahoo2021contamination, cisneros2023combined} and also in the spatial extremes context when censoring non-extreme observations \citep{morris2017space, hazra2018semiparametric, yadav2021spatial, zhang2022hierarchical}}. Furthermore, the spatially-varying parameters and many latent variables allow for Gibbs sampling, providing further computational benefits. 

We fit the proposed model, as well as natural competing models, to analyze heavy rainfall in Bangladesh. We consider daily precipitation data obtained from the Tropical Rainfall Measuring Mission (TRMM Version 7) project, available over the period from March 2000, to December 2019, at a spatial resolution of $0.25^\circ \times 0.25^\circ$. From an agroclimatological perspective, we choose to focus on heavy rainfall that affects monsoon crops, and thus consider data from June to September. Overall, the final dataset has observations at 195 grid cells for 2440 days over the entire country of Bangladesh, which is a relatively large and irregular spatial domain. We demonstrate the performance and usefulness of our modeling approach by drawing posterior inference for return levels at each individual grid cell and also compute regional aggregated risk over six sub-regions defined by \cite{mannan2007climatic}.

The paper is organized as follows. In \S\ref{methodology}, we begin with a summary of the HOT model proposed by \cite{huser2017bridging}, and we then describe our proposed SHOT model, discussing some of its interesting theoretical properties. Bayesian computational details are presented in \S\ref{computation}, with \S\ref{mcmc_sampling} detailing our MCMC algorithm and \S\ref{simulation} reporting the results of an extensive simulation study. We fit the proposed model to analyze heavy precipitation data in Bangladesh in \S\ref{application}, and we finally conclude in \S\ref{conclusion} with some discussion and perspectives on future research. 

\section{Modeling}
\label{methodology}

\subsection{HOT model: a Gaussian single-scale mixture process}
\label{singlescale}
We first define general Gaussian scale mixture processes and then describe the specific HOT model \citep{huser2017bridging}. A Gaussian scale mixture is a process $W(\cdot)$ constructed as 
\begin{equation} \label{hot_w}
    W(\bm{s}) = R Z(\bm{s}), ~~~~~~~~ \bm{s} \in \mathcal{D} \subset \mathbb{R}^2,
\end{equation}
where $Z(\cdot)$ is a standard Gaussian process {(i.e., with zero mean and unit variance marginally)} with correlation function $\rho(\bm{s}_1, \bm{s}_2)=\textrm{Cor}\{Z(\bm{s}_1),  Z(\bm{s}_2)\}$, and $R\sim F_R$ is a non-negative random variable following the distribution $F_R$ and independent of $Z(\cdot)$. Precise tail properties of processes of the form \eqref{hot_w} have been derived in \citet{huser2017bridging} or \citet{EOW2019} for instance; see also the references therein. The HOT model specifically assumes that the mixing distribution, $F_R$, is a two-parameter family, denoted by $F_{\beta, \gamma}$ with parameters $\beta\in[0,\infty)$ and $\gamma>0$, and defined as
\begin{equation} \label{distr_r}
F_{\beta, \gamma}(r) =
  \begin{cases}
    1 - \exp\lbrace - \gamma (r^\beta - 1)/\beta \rbrace & \text{if $\beta > 0$}, \\
    1 - r^{-\gamma} & \text{if $\beta = 0$,}
  \end{cases}
  \qquad r \in [1, \infty).
\end{equation}
This specific family of distributions, $F_{\beta, \gamma}$, is motivated by the different bivariate tail behaviors that the process $W(\cdot)$ in \eqref{hot_w} can capture. Because $F_{\beta, \gamma}$ is Weibull-tailed when $\beta>0$ and Pareto-tailed when $\beta=0$ (or $\beta\to0$), it follows that the vector $\{W(\bm s_1),W(\bm s_2)\}^\top$ can capture both AD (when $\beta=0$) and AI (when $\beta>0$) between any pair of sites $\{\bm s_1,\bm s_2\}\subset \mathcal D$.

To get a finer description of bivariate tail properties, \citet{coles1999dependence} considered the copula-based pair of coefficients $\{\chi,\bar\chi\}$, where $\chi$ is defined as in \eqref{eq:chi} and $\bar\chi=\lim_{u\to1} \bar\chi_u$, with $\bar\chi_u = 2 \log(1-u)/\log\{\bar C(u,u)\} - 1$, where $C(u_1,u_2)$ denotes the copula (i.e., joint distribution with margins transformed to the standard uniform scale) {while the corresponding survival function of the copula $C$ is $\bar{C}(u_1,u_2) = 1 - u_1 - u_2 +~C(u_1, u_2)$.} controlling the random vector under consideration. The $\bar\chi$ coefficient is related to the coefficient of tail dependence introduced by \citet{Ledford1996}, and captures the joint tail decay rate, with slower rates achieved when $\bar\chi$ is larger. Similar to the different asymptotic regimes based on $\chi$, we can show that AD corresponds to $\bar\chi = 1$, and AI to $\bar\chi \in [-1, 1)$, and thus $\bar\chi$ can help identify the asymptotic tail dependence regime and strength by complementing the information provided by $\chi$ itself.

In particular, for the HOT model in \eqref{hot_w}--\eqref{distr_r}, we can show that when $\beta > 0$, we get AI with 
$$\chi = 0~~~ \textrm{and}~~~ \bar \chi = 2 [\{1 + \rho(\bm{s}_1, \bm{s}_2)\}/2]^{\beta / (\beta + 2)} - 1,$$
whereas when $\beta = 0$, we get AD with 
$$\chi = 2 \bar{F}_T \left( [(\gamma + 1)\{1 - \rho(\bm{s}_1, \bm{s}_2)\}/\{1 + \rho(\bm{s}_1, \bm{s}_2)\}]^{1/2}; \gamma + 1 \right)~~~ \textrm{and}~~~ \bar \chi = 1,$$ 
where $F_T(\cdot~; a)$ is the CDF of Student's $t$-distribution with $a$ degrees of freedom, and the corresponding survival function is $\bar F_T(\cdot~; a)=1-F_T(\cdot~; a)$. The HOT model can thus flexibly capture a wide range of tail dependence characteristics by varying the correlation 
$\rho(\bm s_1,\bm s_2)$ and the model parameters $\beta,\gamma$, as demonstrated by the different $\{\chi,\bar\chi\}$ values that can be attained. Moreover, a main benefit of this model over other related approaches is that there is a smooth transition between the two asymptotic regimes as $\beta \downarrow 0$. However, a drawback of the HOT model is that it is constructed from a single $R$ variable for the whole spatial domain, which implies that the $W(\cdot)$ process in \eqref{hot_w} cannot capture full independence unless $R=r_0>0$ is degenerate, which is a clear practical limitation over large domains. Furthermore, this also means that the entire process is either AD or AI for all pairs of sites, and hence, that the model cannot exhibit two different asymptotic regimes for nearby and distant locations.

\subsection{SHOT model: a sparse Gaussian spatial-scale mixture process}
\label{spatialscale}

\subsubsection{General construction} \label{general_construction}
We now present our flexible spatial extension of the HOT model and describe the general construction of the ``core model'' first. Subsequent sections further detail the specifics of each model component, and we finally summarize the full model specification. Inspired by the construction in \eqref{hot_w}, our proposed SHOT model is built by considering a Gaussian process with a spatially-varying (rather than spatially-constant) random scale process, i.e.,
\begin{equation} \label{spatialscale_model}
    X(\bm{s}) = R(\bm{s}) Z(\bm{s}), ~~~~~~~~ \bm{s} \in \mathcal{D} \subset \mathbb{R}^2,
\end{equation}
where $Z(\cdot)$ is a standard Gaussian process with its underlying correlation structure given by $\rho(\bm{s}_1, \bm{s}_2)=\textrm{Cor}[Z(\bm{s}_1),  Z(\bm{s}_2)]$, $R(\cdot)$ and $Z(\cdot)$ are independent spatial processes, and $R(\bm{s})$ is positive with probability one, for all $\bm{s} \in \mathcal{D}$. Conditional on $R(\cdot)$, the process $X(\cdot)$ is Gaussian, with spatially-varying marginal standard deviation, i.e., $\textrm{SD}[X(\bm{s})\mid R(\bm s)] = R(\bm{s})$, and with the same correlation structure as the process $Z(\cdot)$ itself. When $R(\bm{s}) = r_0$ almost surely for all $\bm{s} \in \mathcal{D}$ for some ${r_0>0}$, then the resulting $X(\cdot)$ process is a stationary Gaussian process, which lacks tail flexibility. Our goal in subsequent sections is to design processes $R(\cdot)$ and $Z(\cdot)$ with sparse probabilistic structures (using low-rank and stochastic partial differential
equation (SPDE)-based approximations, respectively) to ensure fast Bayesian inference in high spatial dimensions, and such that the resulting $X(\cdot)$ process enjoys high flexibility in its upper joint tail while remaining approximately tail-stationary, i.e., with ${\rm Pr}(X(\bm{s}) > x)$ being independent of $\bm{s}$ for large $x$, and the corresponding $\chi$-coefficient $\chi_X(\bm{s}_1, \bm{s}_2)$ being a function of the spatial lag $\bm s_1-\bm s_2$ only (in a limiting sense to be made precise below). We note that $W(\cdot)$ in \eqref{hot_w} is trivially tail-stationary assuming $\rho(\bm{s}_1, \bm{s}_2)$ is stationary, while this is less clear for $X(\cdot)$ in \eqref{spatialscale_model}. 

We further note that our proposed construction in \eqref{spatialscale_model} bears resemblance with the non-Gaussian Type-G models of \cite{Bolin.Wallin:2020} and the volatility-modulated moving average model for spatial heteroskedasticity proposed by \cite{Nguyen.Veraart:2017}. The main difference is that the scale variable $R$ in Type-G models is an integral part of the SPDE white noise term rather than externally multiplying the Gaussian SPDE like in our proposed model \eqref{spatialscale_model}. The benefit of decoupling the Gaussian SPDE with the scale process $R(\cdot)$ is that we can get a finer control over the overall extremal dependence structure by modulating the ``mesh size'' of the $R$ process separately from that of the $Z$ process. It is also interesting to remark that the single-scale HOT model with SPDE-based construction for $Z(\cdot)$ is a special Type $G$ model (of Type $G_1$), as defined in \cite{Bolin.Wallin:2020}. 

\subsubsection{Construction of $Z(\cdot)$} \label{z_construction}
We assume throughout that the Gaussian process $Z(\cdot)$ in \eqref{spatialscale_model} has an isotropic Mat\'ern spatial correlation with a nugget effect, i.e., 
\begin{eqnarray} 
\label{cov_structure}
&& \rho(\bm{s}_1, \bm{s}_2) = \ffrac{r}{\Gamma(\nu) 2^{\nu - 1}} \left\{ \ffrac{ d(\bm{s}_1, \bm{s}_2) }{\psi} \right\}^{\nu} K_{\nu} \left\{ \ffrac{ d(\bm{s}_1, \bm{s}_2) }{\psi}  \right\} + (1 - r) \mathbb{I}(\bm{s}_1 = \bm{s}_2),
\end{eqnarray}
where $d(\bm{s}_1, \bm{s}_2)$ is the Euclidean distance between sites $\bm{s}_1$ and $\bm{s}_2$, $\psi > 0$ and $\nu > 0$ are range and smoothness parameters, respectively, $r \in [0, 1]$ is the ratio of spatial to total variation, $K_{\nu}$ is the modified Bessel function of degree $\nu$, and $\mathbb{I}(\cdot)$ is the indicator function. Integer values of $\nu$ determine the degree of mean-square differentiability of $Z(\cdot)$ when $r=1$. However, because $\nu$ is usually hard to identify in practice, it is often fixed in data applications. Here, we choose to fix $\nu=1$, {but in the Supplementary Material we also explore the case $\nu=~2$}. To ensure fast computations, instead of using the exact Mat\'ern Gaussian process with correlation based on \eqref{cov_structure}, which leads to a dense correlation matrix, we approximate $Z(\cdot)$ with an alternative Gaussian process $\tilde{Z}(\cdot)$ built from a Gaussian Markov random field (GMRF) defined on a finite mesh, by exploiting the one-to-one link between continuous-space Mat\'ern Gaussian processes with dense covariance matrices and GMRFs with sparse precision (i.e., inverse covariance) matrices; see \cite{lindgren2011explicit}. This yields the approximate scale mixture process $\tilde{X}(\bm s)=~R(\bm s)\tilde{Z}(\bm s)$, $\bm s \in \calD$. We now briefly review this approach.

\begin{figure}[t!]
\centering
\adjincludegraphics[height = 0.4\linewidth , trim = {{.0\width} {.0\width} {.0\width} {.0\width}}, clip]{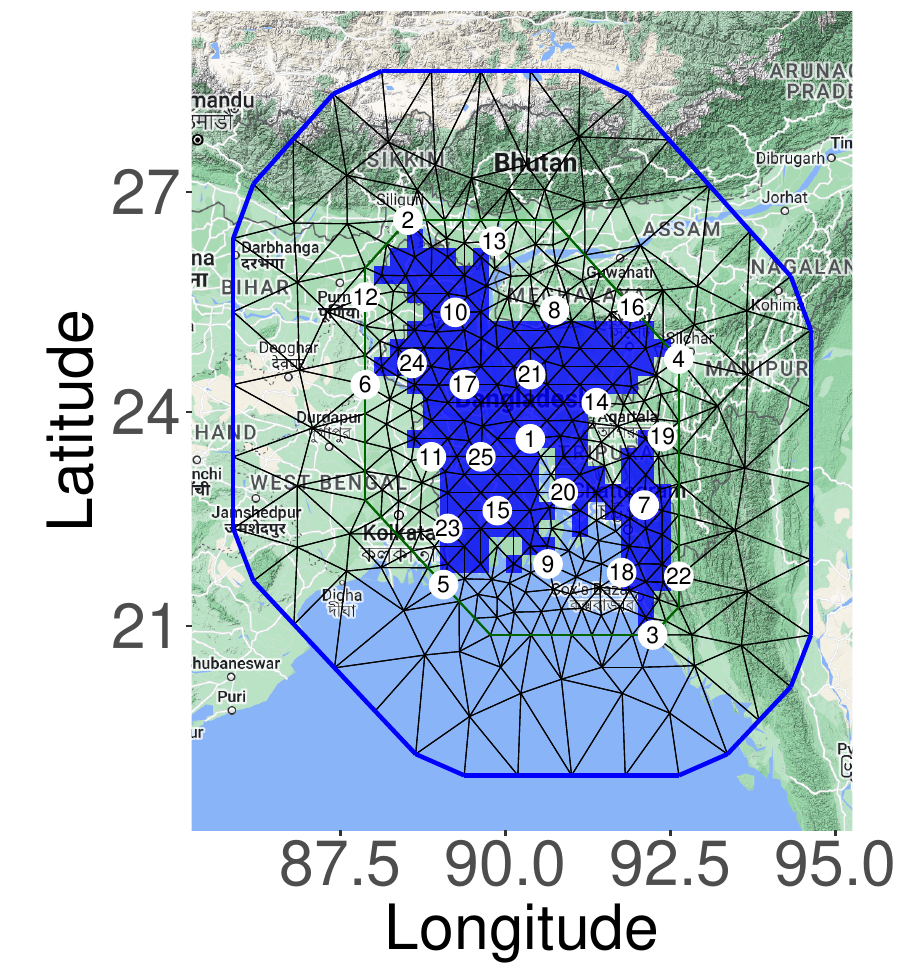} 
\adjincludegraphics[height = 0.4\linewidth, trim = {{.0\width} {.0\width} 0 {.0\width}}, clip]{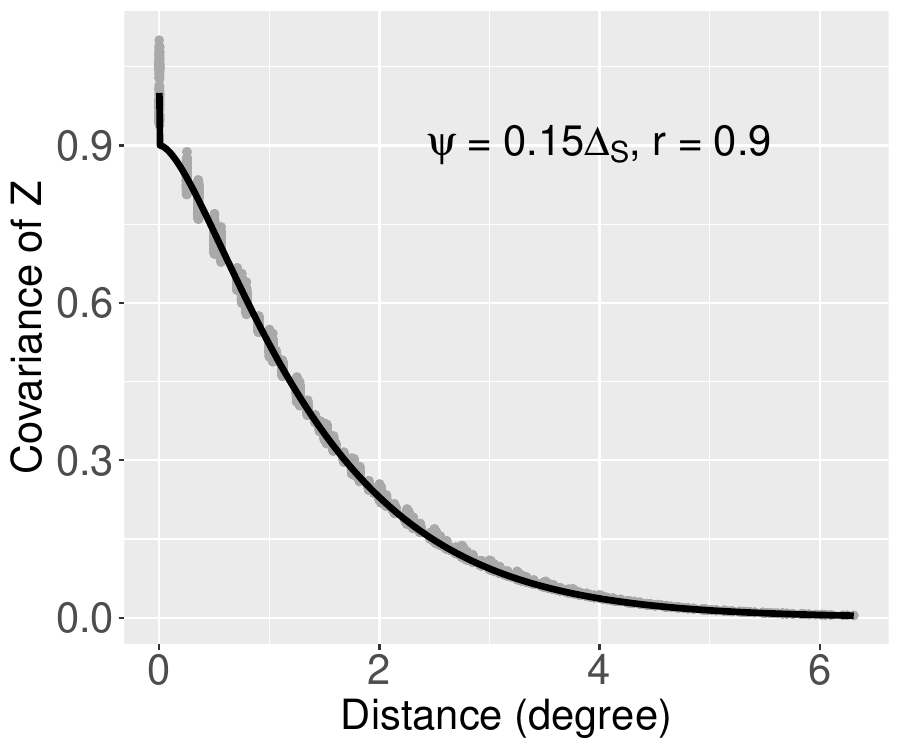}
\vspace{-2mm}
\caption{\emph{Left:} Triangulated mesh over Bangladesh, that is used to approximate the spatial SPDE process $Z(\cdot)$. The {numbered} dots denote the spatial knot locations $\mathcal{S}^*$ used in the construction of the random scale process $R(\cdot)$. \emph{Right:} Comparison of the true Mat\'ern correlation ({ solid line}) and the pairwise covariances between two spatial locations obtained by the SPDE approximation ({points}), as a function of distance. The parameters are set to $\psi = 0.15\Delta_{\mathcal{S}}$, with $\Delta_{\mathcal{S}}$ the maximum spatial distance between two locations in the domain, and $r = 0.9$.}
\label{spde_approximation}
\end{figure}

Suppose that $\varepsilon(\cdot)$ is a Gaussian process with spatial correlation structure \eqref{cov_structure} and $r=1$. Then, for $\nu=1$, $\varepsilon(\cdot)$ is the solution to the linear SPDE
\begin{equation}
    \label{eq:spde.fem}
    (\psi^{-2} - \Delta) \varepsilon(\bm{s}) = 4\pi \psi^{-2}\mathcal{W}(\bm{s}),\qquad \bm{s}\in\mathbb R^2,
\end{equation}
where $\mathcal{W}(\bm{s})$ is Gaussian white noise, and $\Delta$ 
is the Laplacian. The solution $\varepsilon(\bm{s})$ to the SPDE may then be approximated using finite-element methods \citep{bolin2024covariance} over a triangulated mesh defined on a bounded domain in $\mathbb{R}^2$, where the triangles are formed as a refined Delaunay triangulation. In \texttt{R}, the mesh can be easily constructed using the function \texttt{inla.mesh.2d} implemented in the \texttt{R} package \texttt{INLA} (\url{www.r-inla.org}), see \citet{Lindgren.Rue:2015}. The left panel of Figure~\ref{spde_approximation} illustrates the mesh that we consider in the data application in Section~\ref{application}.  
Suppose now that the data are observed at the spatial locations $\mathcal{S}=\{\bm{s}_1, \ldots, \bm{s}_N\}\subset\mathcal D$, and let $\mathcal{S}^* = \{ \bm{s}^*_1, \ldots, \bm{s}^*_{N^*} \}$ be the set of mesh nodes. We construct a finite-element representation of the solution by writing $\tilde{\varepsilon}(\bm{s}) = \sum_{j=1}^{N^*} \zeta_j(\bm{s}) \varepsilon^*_j$ and plugging it into $\varepsilon(\cdot)$ in \eqref{eq:spde.fem}, where $\{\zeta_j(\cdot)\}$ are piecewise linear and compactly-supported ``hat'' basis functions defined on the mesh and $\{\varepsilon^*_j\}$ are normally distributed weights defined for each basis function (i.e., one for each mesh node in $\mathcal{S}^*$). Using suitable test functions, we can then show that the vector $\bm{\varepsilon}^* = [\varepsilon^*_1, \ldots, \varepsilon^*_{N^*}]'$ has distribution $\bm{\varepsilon}^* \sim \textrm{Normal}_{N^*}(\bm{0}, \bm{Q}_{\psi}^{-1})$, with precision matrix $\bm{Q}_{\psi}$ given in explicit form by $\bm{Q}_{\psi} = 0.25 \pi^{-1} \left[ \psi^{-2} \bm{D} +  2\bm{G}_1 + \psi^{2}\bm{G}_2 \right]$,
where $\bm{D}$, $\bm{G}_1$, and $\bm{G}_2$ are \emph{sparse} $(N^* \times N^*)$-dimensional finite-element matrices obtained as follows. Defining the inner product $\langle\, f, g \rangle = \int f(\bm{s}) g(\bm{s}) {\rm d}\bm{s}$, the matrix $\bm{D}$ is diagonal with $(j,j)$th entry $D_{j,j} = \langle\, \zeta_j(\cdot), 1 \rangle$; the matrix $\bm{G}_1$ has $(j_1,j_2)$th entry $G_{1;j_1,j_2} = \langle\, \nabla \zeta_{j_1}(\cdot), \nabla \zeta_{j_2}(\cdot) \rangle$; and $\bm{G}_2 = \bm{G}_1 \bm{D}^{-1} \bm{G}_1$. Efficient computation of these sparse matrices is implemented in the function \texttt{inla.mesh.fem} from the \texttt{R} package \texttt{INLA}. See also \cite{bakka2018spatial} and \citet{Lindgren.etal:2021} for further theoretical details.

To project the spatial random effects $\bm{\varepsilon}^*$ (defined over the mesh nodes) back to the observation locations in $\mathcal{S}$, we define an $(N \times N^*)$-dimensional SPDE projection matrix $\bm{A}$, whose $(i,j)$th entry is equal to $\zeta_j(\bm{s}_i)$ for each spatial location $\bm{s}_i\in\mathcal S$ and the mesh node $\bm{s}_j^*\in\mathcal S^*$, and then compute $\bm{A}\bm{\varepsilon}^*$. The construction of the matrix $\bm{A}$ is implemented in the function \texttt{inla.spde.make.A} from the \texttt{R} package \texttt{INLA}. When there is a nugget effect $r\in[0,1]$ as in \eqref{cov_structure}, our model for the observations $\tilde{\bm{Z}}=[\tilde{Z}(\bm{s}_1), \ldots, \tilde{Z}(\bm{s}_N)]'$ thus becomes
\begin{eqnarray} \label{z_construction_eq}
\tilde{\bm{Z}} = \sqrt{r} \bm{A} \bm{\varepsilon}^* + \sqrt{1-r}\, \bm{\eta},
\end{eqnarray}
where $\bm{\eta} = [\eta_1, \ldots, \eta_N]'$ with $\eta_i \overset{\textrm{IID}}{\sim} \textrm{Normal}(0, 1)$. From \eqref{z_construction_eq}, the covariance matrix of $\tilde{\bm{Z}}$ can be expressed as $\Sigma_{\tilde{\bm{Z}}} = r\bm{A}\bm{Q}_{\psi}^{-1}\bm{A}' + (1-r)\bm{I}_N$.

To illustrate this approximation in the context of our data application, we consider the domain and the mesh displayed in the left panel of Figure~\ref{spde_approximation}, and set $\psi = 0.15 \Delta_{\mathcal{S}}$ and $r = 0.9$, where $\Delta_{\mathcal{S}}$ denotes the maximum Euclidean distance between any two spatial locations. We then compute the approximate covariance between every pair of spatial locations (i.e., the elements of the matrix $\Sigma_{\tilde{\bm{Z}}}$), as well as the true Mat\'ern correlation based on \eqref{cov_structure}, and plot them in the right panel of Figure~\ref{spde_approximation}, as a function of distance. We can see that the SPDE approach indeed provides an accurate approximation to the true correlation structure. We shall therefore exploit the sparsity of the matrix $\bm{Q}_{\psi}$ for fast Bayesian computations. Additionally, we shall use the conditional independence structure of $\tilde{\bm{Z}} \mid \bm{\varepsilon}^* \sim \textrm{Normal}_{N}(\sqrt{r} \bm{A} \bm{\varepsilon}^*, (1-r)\bm{I}_N)$ for fast (univariate) imputation of non-extreme data treated here as being left-censored.

\subsubsection{Construction of $R(\cdot)$} \label{r_construction}

For the process $R(\cdot)$ in \eqref{spatialscale_model}, we consider the low-rank representation 
\begin{equation} \label{lowrank}
R(\bm{s}) = \begin{cases}
    \left(1 + \beta \log\left[ \sum_{k=1}^K \{B_k(\bm{s}; \phi)\}^{1 / \gamma} \exp\left(\ffrac{R_k^{*\beta} - 1}{\beta} \right) \right] \right)^{1 / \beta} & \text{if $\beta > 0$}, \\
    \sum_{k=1}^K \{B_k(\bm{s}; \phi)\}^{1/\gamma} R^*_k & \text{if $\beta = 0$,}
  \end{cases}
  \qquad \gamma>0,
\end{equation}
where $\bm{R}^*=[R_1^*,\ldots,R_K^*]'$ are independent latent random effects distributed according to $F_{\beta, \gamma}$ defined as in the HOT model in \eqref{distr_r}, i.e., $R^*_k \overset{\textrm{IID}}{\sim} F_{\beta, \gamma}$, and $\{B_k(\cdot; \phi)\}$ are (fixed) compactly-supported spatial basis functions satisfying the sum-to-one constraint $\sum_{k=1}^{K} B_k(\bm{s}; \phi) = 1$ for each $\bm{s} \in \mathcal{S}$. Note that when $K=1$ in \eqref{lowrank}, the spatial-scale mixture model (\ref{spatialscale_model}) coincides with the single-scale mixture model \eqref{hot_w}, and that the process for $\beta = 0$ can also be obtained as the limit as $\beta \downarrow 0$. While the process built from $\beta=0$ may appear to have some vague resemblance with the \citet{reich2012hierarchical} max-stable process, we emphasize that our proposed model is not max-stable and, as we shall see in \S\ref{model_props}, it has a considerably more flexible dependence structure that is not restricted to be AD either. The choice of $K$ is problem-specific and involves a trade-off between the approximation to a stationary process and the computational burden. Intuitively, as $K$ becomes larger and the basis functions become more densely spaced over the whole domain, the value of the process $R(\cdot)$ at a given site thus becomes influenced by the contributions of increasingly many basis functions, whose exact location thus becomes ``irrelevant'', leading to stationarity. The construction obtained as $K\to\infty$ is similar to a certain form of kernel convolution \citep{Krupskii.Huser:2021}, although in the case of the process \eqref{lowrank}, the variables $R_k^*$ are not infinitely divisible and so, the limit does not have a pointwise meaning; nevertheless, the process $R(\cdot)$ in \eqref{lowrank} still becomes approximately stationary as $K\to\infty$. However, a large $K$ also implies a larger number of unobserved latent factors $R_k^*$, which makes it computationally more challenging; see \S\ref{model_props} for more details. In our simulations and data application, we consider a few different choices for $K$ and then select the most optimal one based on the {deviance information criterion (DIC)}. 

\begin{figure}[t!]
\centering
	\adjincludegraphics[height = 0.31\linewidth, trim = {{.0\width} {.0\width} {.0\width} {.0\width}}, clip]{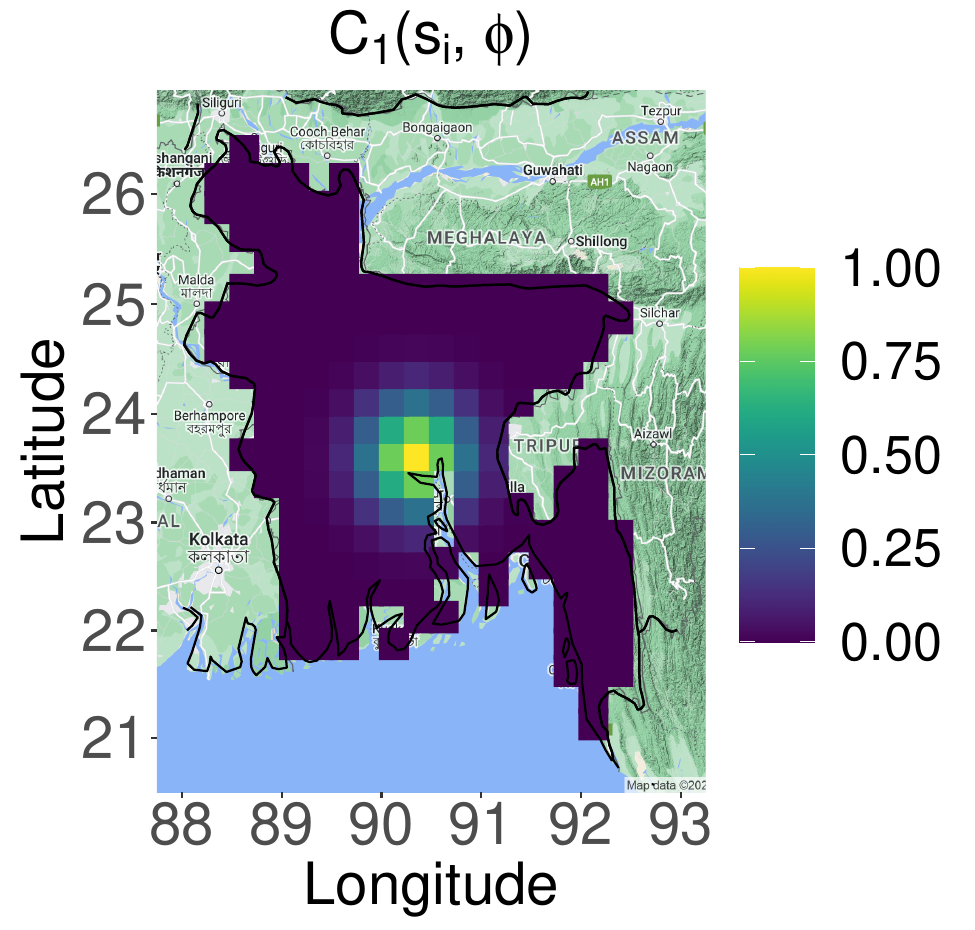}
	\adjincludegraphics[height = 0.31\linewidth, trim = {{.0\width} {.0\width} {.0\width} {.0\width}}, clip]{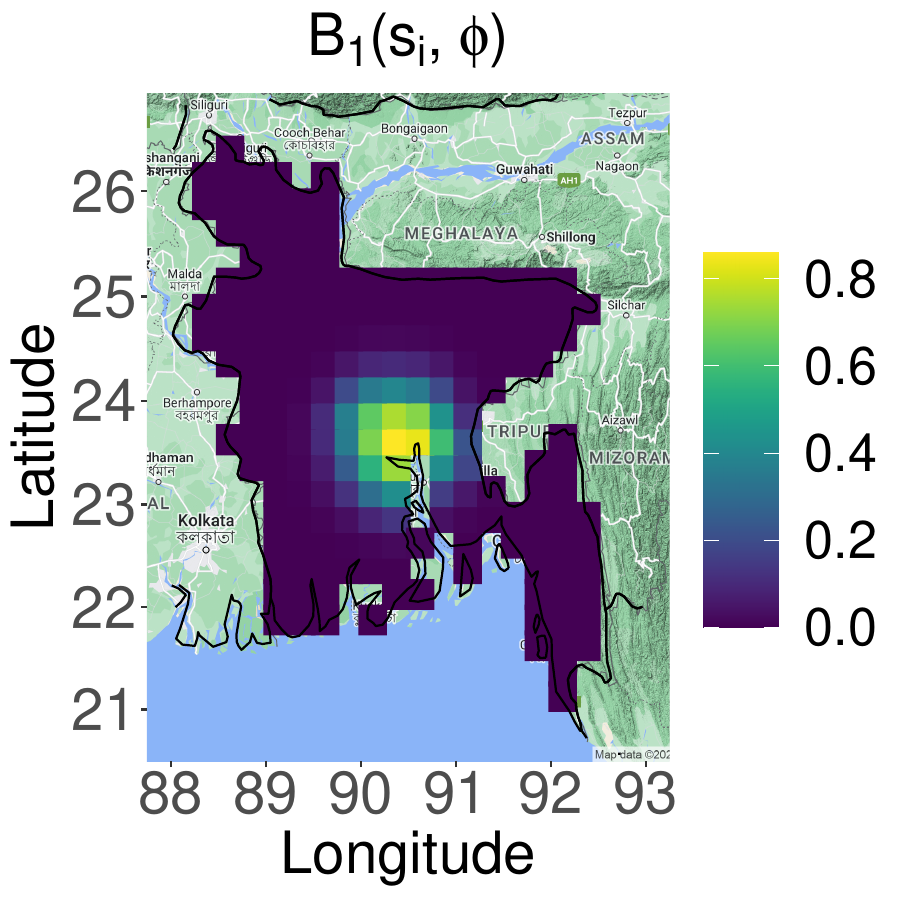}
	\adjincludegraphics[height = 0.31\linewidth, trim = {{.0\width} {.0\width} {.0\width} {.0\width}}, clip]{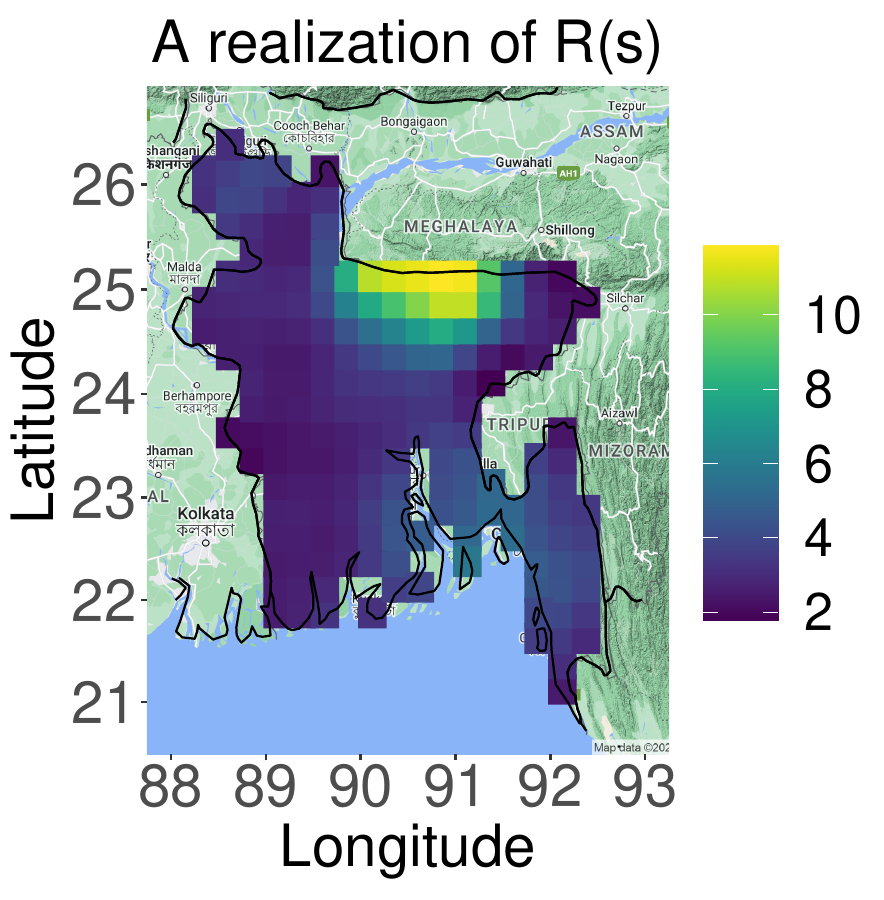}
	\caption{\emph{Left:} Wendland basis function $C_{1}(\bm{s}_i; \phi)$ evaluated at the data locations $\bm s_i\in\mathcal{S}$. \emph{Middle:} The corresponding rescaled Wendland basis function $B_{1}(\bm{s}_i; \phi)$. \emph{Right:} A simulated realization of the $R(\cdot)$  process for $\beta = 0$, $\gamma = 2.5$, $\phi = 1.5$, and $K=25$ in \eqref{lowrank}.}
	\label{scale_construction}
\end{figure}

We now discuss the construction of the basis functions $\{B_k(\cdot; \phi)\}$, whose purpose and definition differ from the ``hat'' basis functions used to build the finite-element approximation to the SPDE solution \eqref{eq:spde.fem}. To allow the stochastic process (\ref{spatialscale_model}) to have extremal dependence only at short distances, while making sure the $R(\cdot)$ process varies smoothly over space, we choose compactly-supported Wendland basis functions of order 2 over $\mathbb{R}^2$ (implemented in the function \texttt{Wendland2.2} from the \texttt{R} package \texttt{fields} {\citet{fields:2021}}), i.e.,
\begin{eqnarray} \label{basis_wendland}
C_k(\bm{s}; \phi) = \left\{1 - \ffrac{d(\bm{s},\tilde{\bm{s}}_k)}{\phi}\right\}^6 \left\{35 \ffrac{d(\bm{s},\tilde{\bm{s}}_k)^2}{\phi^2} + 18 \ffrac{d(\bm{s},\tilde{\bm{s}}_k)}{\phi} + 3 \right\} \mathbb{I}\{d(\bm{s},\tilde{\bm{s}}_k) < \phi\},
\end{eqnarray}
where $\phi>0$ is a range parameter, and $\tilde{\mathcal{S}}= \{\tilde{\bm{s}}_1, \ldots, \tilde{\bm{s}}_K \}\subset \mathcal D$ denotes a set of $K$ spatial knot locations; we then set $B_k(\bm{s}; \phi) = C_k(\bm{s}; \phi) / \sum_{k'=1}^K C_{k'}(\bm{s}; \phi)$, $k=1,\ldots,K$, to satisfy the sum-to-one constraint. Wendland basis functions have been quite popular in spatial statistics to construct multiresolution Gaussian processes \citep{Nychka.etal:2015}, due to their compact support, optimality properties, and ability to capture different degrees of smoothness. With an order of 2, we here get moderately smooth basis functions. While $\tilde{\mathcal{S}}$ does not necessarily need to coincide with any of the mesh nodes $\mathcal S^*$ used to construct the GMRF $\tilde{\bm{Z}}$ in \eqref{eq:spde.fem}--\eqref{z_construction_eq}, we still choose $\tilde{\mathcal{S}}\subset\mathcal{S}^*=\{\bm{s}^*_1, \ldots, \bm{s}^*_{N^*} \}$, with $K\ll N^*$. 
Precisely, we select the knot locations as the first $K$ locations from $\mathcal{S}^{*(c)} = \cup_{\{ \bm{s}_i \in \mathcal{S} \}} \{\bm{s}^*_{j} \in \mathcal{S}^* \mid d(\bm{s}_i, \bm{s}^*_{j} ) \leq c \cdot \max_{\bm{s}^*_{j'} \in \mathcal{S}^*} d(\bm{s}_i, \bm{s}^*_{j'}) \} \subseteq \mathcal{S}^*$ (i.e., the set of mesh nodes that are within a certain distance from the data locations), according to the maximum-minimum ordering introduced by \cite{guinness2018permutation} in the context of the Vecchia Gaussian likelihood approximation. When $c=1$, we retrieve $\mathcal{S}^{*(c)} = \mathcal{S}^*$, but we set $c=0.05$ in our simulations and data application; this avoids selecting knot locations that are too far away from the observation locations to ensure approximate stationarity (see \S\ref{model_props}). The maximum-minimum ordering is a space-filling design, which spreads out knot locations to cover the space $\mathcal S^{*(c)}$ as evenly as possible. Specifically, the initial knot location is selected as the mesh node that has the least average distance to the other mesh nodes (i.e., it is the most central). Subsequently, the $j$th mesh node is selected from $\mathcal{S}^{*(c)}$ as the $k$th knot location if
\begin{eqnarray}
\nonumber j = \argmax_{\{j' \mid \bm{s}^*_{j'} \in \mathcal{S}^{*(c)} \setminus \{\tilde{\bm{s}}_1, \ldots, \tilde{\bm{s}}_{k-1} \} \} } \min_{k' \in \{1,\ldots,k-1 \}}d(\bm{s}^*_{j'}, \tilde{\bm{s}}_{k'}).
\end{eqnarray}
The left panel of Figure~\ref{spde_approximation} illustrates this ordering with $K=25$ and $c=0.05$, while Figure~\ref{scale_construction} displays the first Wendland basis function $C_1(\bm{s}_i; \phi)$, its rescaled version $B_1(\bm{s}_i; \phi)$, and a realization of $R(\cdot)$ for $\phi = 1.5$ at all observation locations in $\mathcal{S}$. 
The compactness of $B_{1}(\bm{s}_i; \phi)$ appears clearly from the middle panel, and hence its corresponding random coefficient impacts only the observations in the central region, thereby allowing for local shocks as desired. The realization of $R(\cdot)$ shows that the proposed construction indeed produces a spatially-varying scale process instead of a single, spatially-constant variable $R$ throughout the domain as in \eqref{hot_w}. In other words, our SHOT model has the ability to realistically generate large $R$ values in a localized region only, e.g., the Himalayan foothills, whereas the HOT model cannot.

\subsubsection{Full model specification} 

The core model \eqref{spatialscale_model} is meant to capture a wide range of extremal dependence structures, and to achieve (approximate) tail-stationarity; see \S\ref{model_props} for more details on further properties. {However, while the assumption of a stationary extremal dependence structure is reasonable for real data on moderately large domains, the marginal distributions often exhibit nonstationarity.} To allow for additional flexibility in the model's marginal tails, we modify the model as 
\begin{eqnarray} \label{full_model}
Y(\bm{s}) = \mu(\bm{s}) + \tau^{-1/2} X(\bm{s}),\qquad \bm s\in\mathcal S\subset\mathcal D,
\end{eqnarray}
where the $X(\cdot)$ process is constructed as in \eqref{spatialscale_model}, $\tau>0$ is a spatially-constant precision parameter, and $\mu(\cdot)$ is a spatially-varying location surface modeled through covariates. Note that in our data application, we standardize the data in a preprocessing step to stabilize inference, so it is not necessary to let the precision parameter $\tau$ be spatially-varying. Moreover, the full model \eqref{full_model} is only assumed to fit well extreme observations beyond a certain high threshold. In other words, considering the data to be time-independent realizations $\mathcal{Y} = \{Y_t(\bm{s}); \bm{s} \in \mathcal{S}, t=1,\ldots,T \}$ of the process $Y(\bm s)$ in \eqref{full_model}, the spatial model \eqref{full_model} is assumed to accurately describe both marginal and dependence \emph{tail} characteristics. To calibrate the model in its upper tail region, we choose site-specific thresholds (e.g., the $95$th percentile at each site) and fit the model to threshold exceedances, treating the values falling below their corresponding threshold as left-censored. In our Bayesian algorithm, we impute these censored values using an efficient method that exploits conditional independence at the data level; see \S\ref{computation} and the Supplementary Material for more details. 

\subsection{Model properties}
\label{model_props}
We here study the tail properties of the core model \eqref{spatialscale_model}. While the process $X(\cdot)$ becomes approximately stationary as $K\to\infty$ (recall the discussion in Section~\ref{r_construction}), we now show that the margins of our model are approximately \emph{tail-stationary} for any \emph{finite} $K$.

\begin{prop} \label{thm_tail_stationary}
For any $\beta\geq 0$ and $K\geq1$, the process $R(\cdot)$ has marginal tails given by 
$${\rm Pr}\{R(\bm{s}) > r\}\sim 1-F_{\beta,\gamma}(r),\quad r\to\infty,~~\bm{s} \in \mathcal{S},$$
where $F_{\beta,\gamma}$ is defined in \eqref{distr_r}.
\end{prop}
A formal proof is provided in the Supplementary Material. A direct corollary of Proposition~\ref{thm_tail_stationary} implies that $\lim_{r\uparrow \infty} {\rm Pr}\{R(\bm{s}_1) > r\} \big/ {{\rm Pr}\{R(\bm{s}_2) > r\}}  = 1$ for all $\bm{s}_1, \bm{s}_2 \in \mathcal{S}$, i.e., $R(\bm{s})$ indeed satisfies marginal tail-stationarity. Moreover, this relation is also valid for the spatial stochastic process $X(\cdot)$ defined in \eqref{spatialscale_model}, i.e., $\lim_{x\uparrow \infty}{\rm Pr}\{X(\bm{s}_1) > x\} / {\rm Pr}\{X(\bm{s}_2) > x\} = 1$ for all $\bm{s}_1, \bm{s}_2 \in \mathcal{S}$, since $X(\bm{s}) = R(\bm{s})Z(\bm{s})$ with $Z(\bm{s}) \sim \textrm{Normal}(0,1)$, while this holds only approximately for $\tilde{X}(\bm{s}) = R(\bm{s})\tilde{Z}(\bm{s})$ when the GMRF approximation is used (recall \S\ref{z_construction}). Moreover, as the tail of $R(\bm s)$ is driven by $F_{\beta,\gamma}$, the Gaussian scale mixture $X(\bm s)$ is (heavy) Pareto-tailed with tail index $1/\gamma>0$ when $\beta=0$ and Weibull-tailed with index $2\beta/(2+\beta)$ when $\beta>0$; see \citet{breiman1965some}, \citet{Arendarczyk.Debicki:2011} and \citet{huser2017bridging} for details. {We note that our model can thus capture light (for $\beta\geq2$) to heavy (for $\beta<2$) tails marginally, with power-law tails arising when $\beta=0$; this possibility to capture heavy tails is an important consideration for our data application in Section~\ref{application}, given that this behavior is consistent with what is typically observed in rainfall data.}

We now discuss the tail dependence properties of the scale process $R(\cdot)$ and the core model $X(\cdot)$. Our main focus in on the tail-correlation coefficient $\chi$ defined in \eqref{eq:chi} when $\beta=0$, { but further results on the sub-asymptotic behavior in terms of the $\bar\chi$ coefficient (recall Section~\ref{singlescale}) are provided in the Supplementary Material for $\beta=0$}. We add subscripts, using obvious notation, to indicate the process under study. When $\beta > 0$, we have $\chi_X(\bm s_1,\bm s_2)=0$, which follows from \cite{huser2017bridging} and the fact that the SHOT process $X(\cdot)$ in \eqref{spatialscale_model} exhibits weaker tail dependence than that of the HOT process $W(\cdot)$ in \eqref{hot_w}; {deriving the $\bar\chi$ coefficient of the SHOT model when $\beta>0$ remains an open question.} 

\begin{prop}\label{thm_tail_dependence}
When $\beta=0$, the tail-correlation coefficient for the $R(\cdot)$ process in \eqref{lowrank} is
    \begin{eqnarray}
        \nonumber \chi_{R}(\bm{s}_1, \bm{s}_2) = \sum_{k=1}^K \min\{ B_k(\bm{s}_1; \phi), B_k(\bm{s}_2; \phi)\}.
    \end{eqnarray}
\end{prop}

\begin{figure}[t!]
\centering
 \adjincludegraphics[height = 0.3\linewidth, trim = {{.0\width} {.0\width} {.0\width} {.0\width}}, clip]{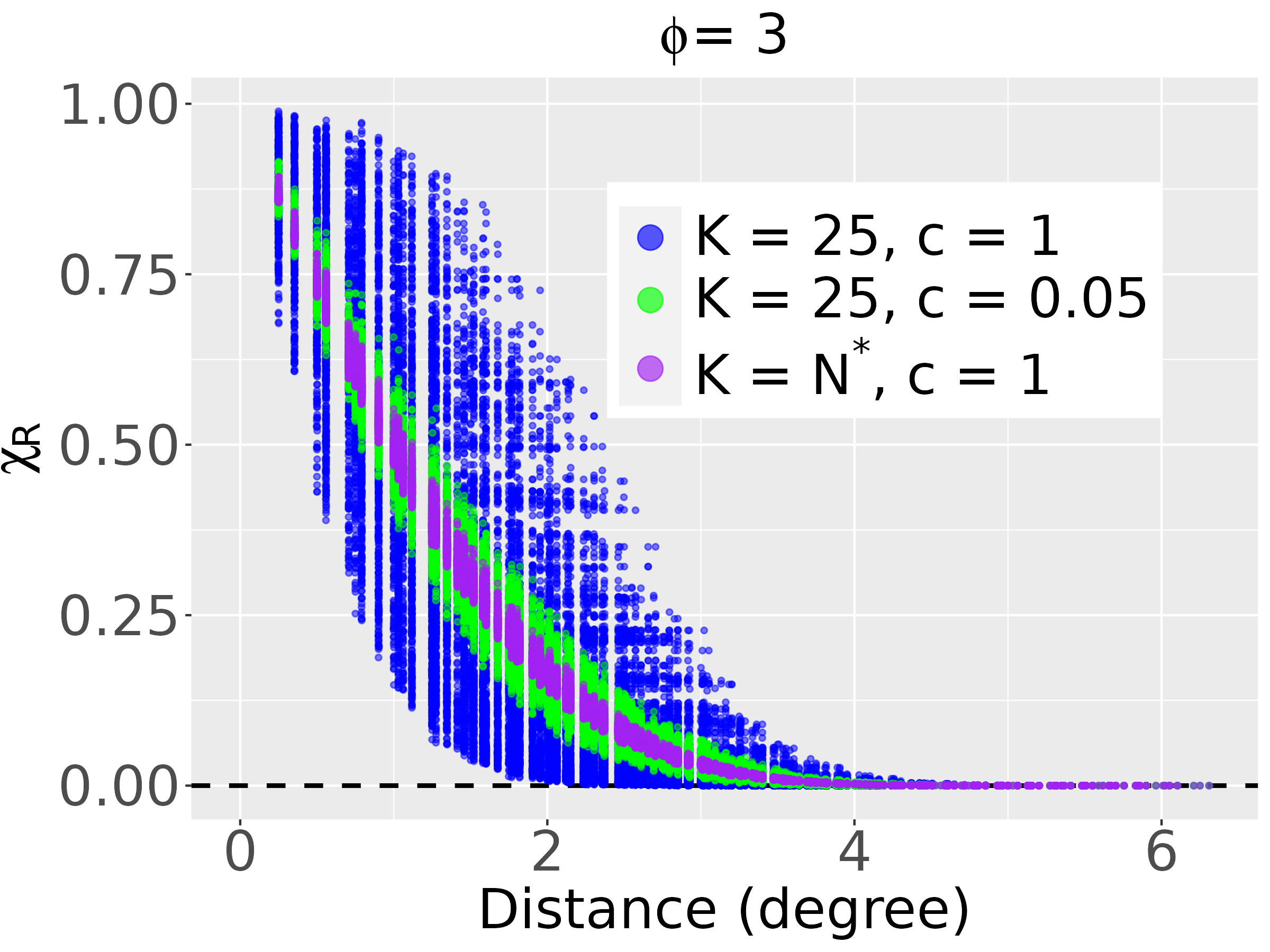}
	\adjincludegraphics[height = 0.3\linewidth, trim = {{.0\width} {.0\width} {.0\width} {.0\width}}, clip]{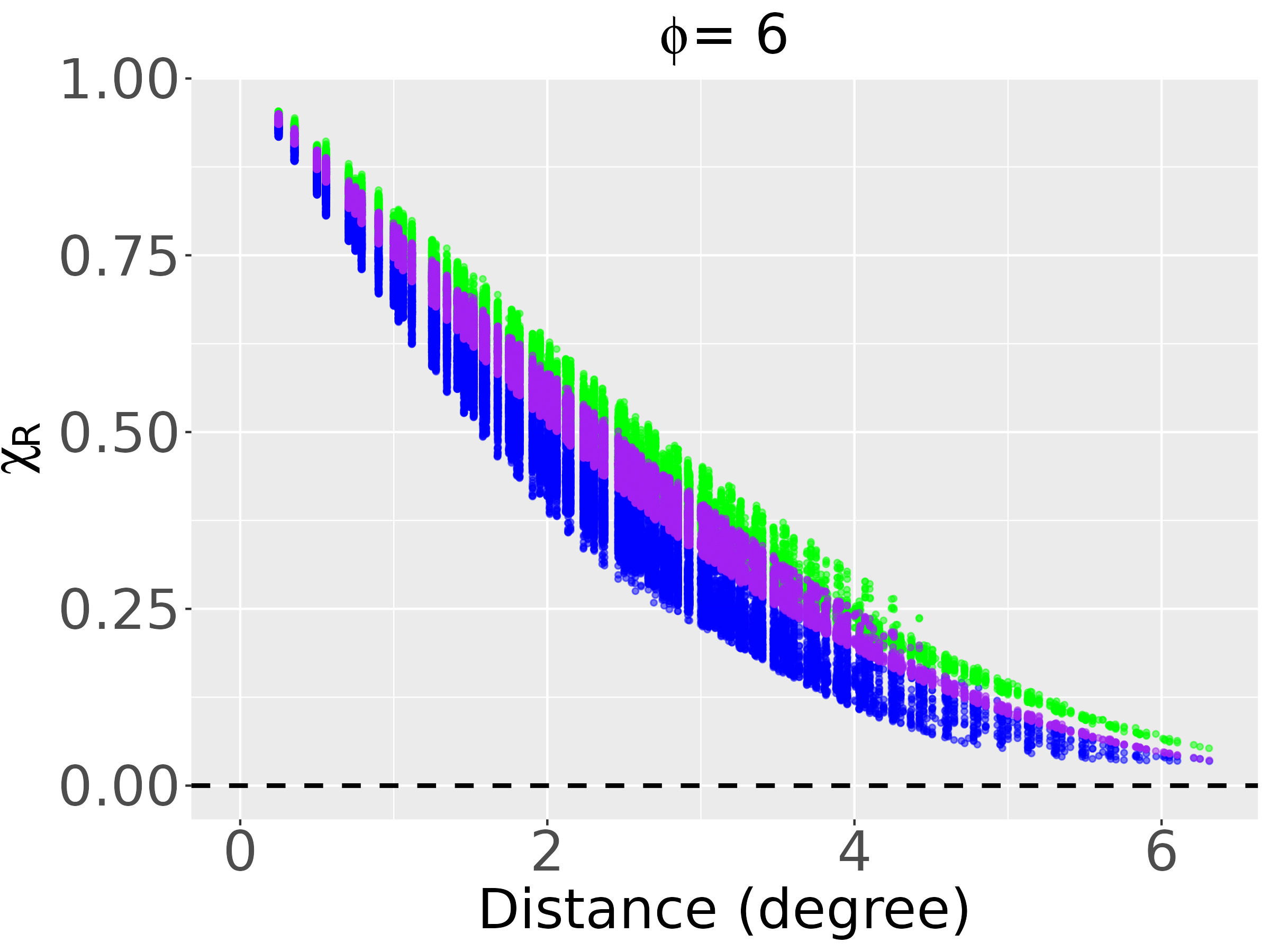}
	\adjincludegraphics[height = 0.3\linewidth, trim = {{.0\width} {.0\width} {.0\width} {.0\width}}, clip]{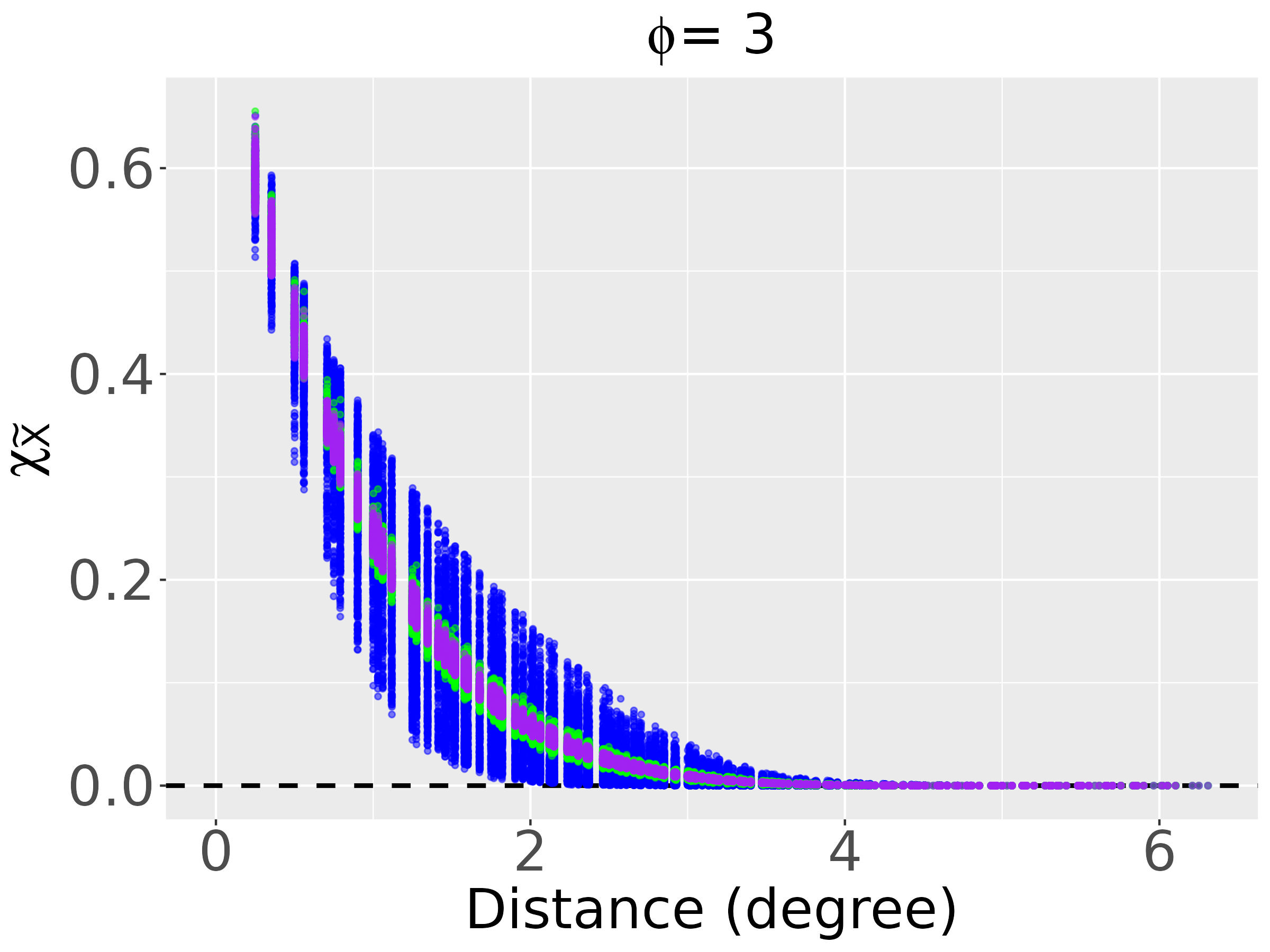}
	\adjincludegraphics[height = 0.3\linewidth, trim = {{.0\width} {.0\width} {.0\width} {.0\width}}, clip]{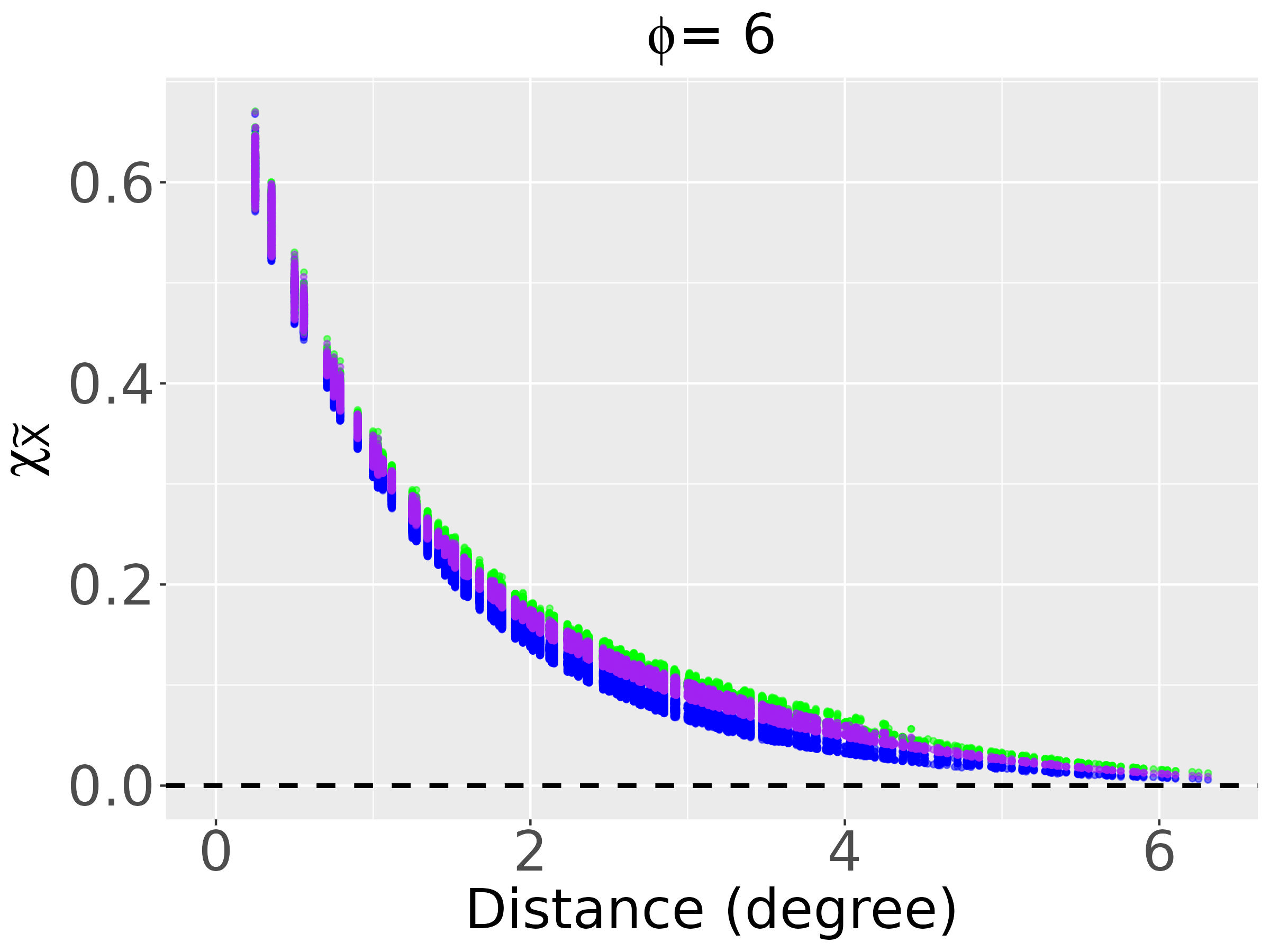}
	\caption{{Tail correlation coefficient $\chi_{R}(\bm s_1,\bm s_2)$ (top) and $\chi_{\tilde{X}}(\bm s_1,\bm s_2)$ (bottom) of the random scale process $R(\cdot)$} and the GMRF-based mixture process $\tilde{X}(\cdot)$, respectively, plotted as functions of distance $\|\bm s_1-\bm s_2\|$, based on Wendland basis functions with range parameter $\phi=3$ degrees (left) and $\phi=6$ degrees (right) in \eqref{basis_wendland}, for the locations/mesh shown in Figure~\ref{spde_approximation} under three settings: (i) $K = 25$, $c=1$; (ii) $K = N^*$, $c=1$; and (iii) $K = 25$, $c=0.05$. We set $\psi = 0.15\Delta_{\mathcal{S}}$, $r=0.9$, and $\nu=1$ in \eqref{cov_structure} and $\beta=0,\gamma = 2.5$ in \eqref{lowrank}. All subfigures share the same legend.}
	\label{chi_X_illustration}
\end{figure}

A formal proof is provided in the Supplementary Material. Interestingly, the tail correlation coefficient $\chi_{R}(\bm{s}_1, \bm{s}_2)$ is thus independent of $\gamma$. Moreover, we have that $\chi_R(\bm{s}, \bm{s})=1$ as expected, and if there is no basis function ``covering'' both locations $\bm{s}_1$ and $\bm{s}_2$ simultaneously, then $\chi_R(\bm{s}_1,\bm{s}_2)=0$; {this is illustrated in the top panels of Figure \ref{chi_X_illustration}}. 
Because we consider compactly-supported basis functions $B_k(\cdot, \phi)$, {the parameter $\phi$ thus controls the strength as well as the range of spatial extremal dependence.} To get further insight, assume that the knot locations $\tilde{\mathcal S}=\{\tilde{\bm s}_1,\ldots,\tilde{\bm s}_K\}$ have been drawn from a probability measure $\nu$ over $\mathcal D=\mathbb R^2$. Then, after integrating out $\tilde{\mathcal S}$, we get 
\begin{eqnarray}
\nonumber  \chi_{R}(\bm{s}_1, \bm{s}_2) &=& \sum_{k=1}^K\E[\min\{ B_k(\bm{s}_1; \phi), B_k(\bm{s}_2; \phi)\}]=K\E[\min\{ B_1(\bm{s}_1; \phi), B_1(\bm{s}_2; \phi)\}] \\ 
\nonumber  & \approx & \int_{\mathcal D}\min\{ C_1(\bm{s}_1-\bm{t}; \phi), C_1(\bm{s}_2-\bm{t}; \phi)\}\nu({\rm d}\bm{t}) \bigg/ \int_{\mathcal D}C_1(\bm{s}-\bm{t}; \phi)\nu({\rm d}\bm{t})\\
\nonumber  &=& \int_{\mathcal D}\min\{ C_1(\bm{t}; \phi), C_1(\bm{s}_2-\bm{s}_1+\bm{t}; \phi)\}\nu({\rm d}\bm{t}) \bigg/ \int_{\mathcal D}C_1(\bm{t}; \phi)\nu({\rm d}\bm{t})~~\in[0,1],
\end{eqnarray}
where the approximation is justified by the law of large numbers when $K\to\infty$. As this expression depends on $\{\bm s_1,\bm s_2\}$ only through the spatial lag $\bm s_2-\bm s_1$, we see again that the $R(\cdot)$ process is approximately tail-stationary as $K\to\infty$.

Practically, $K$ will always be finite, which induces some nonstationarity artifacts. However, this can be mitigated by carefully choosing the knot locations (e.g., using the maximum-minimum design detailed in \S\ref{r_construction}) and an appropriate value of $K$, both of which should also ideally depend on the unknown range parameter $\phi$. Intuitively, if $\phi$ is large compared to the distance between knot locations, then the $R(\cdot)$ process will be nearly stationary, while if it is too small the resulting process might be strongly nonstationary. In addition, since the basis functions are compactly supported, $\phi$ cannot be arbitrarily small; {we must ensure that every location $\bm{s}\in\mathcal S$ is ``covered'' by at least one basis function, so that $\sum_{k'=1}^K C_{k'}(\bm s;\phi)\neq 0$ in order that $B_k(\bm s;\phi)$ remains finite and the process be well-defined.} To satisfy this criterion, we need $\min_{\bm s\in\mathcal S} \max_{k=1}^K C_k(\bm{s}, \phi) > 0$, which is equivalent to imposing $\phi>\phi_{\min}=\max_{\bm s\in\mathcal S}\min_{k=1}^K d(\bm s,\tilde{\bm{s}}_k)$. On the opposite side of the spectrum, when $\max_{k=1}^K \min_{\bm s\in\mathcal S} C_k(\bm{s}, \phi) > 0$, which is equivalent $\phi>\phi_{\max}=\min_{k=1}^K\max_{\bm s\in\mathcal S} d(\bm s,\tilde{\bm{s}}_k)$, then there is at least one basis function that covers \emph{all} sites simultaneously, resulting in strong tail dependence across the domain and $\chi_{R}(\bm{s}_1, \bm{s}_2) > 0$ for any $\bm{s}_1, \bm{s}_2 \in \mathcal{S}$. In practice, we might want to restrict $\phi_{\min}<\phi<\phi_{\max}$, unless there is very strong tail dependence in the data.

For the mesh and the knot locations shown in the left panel of Figure~\ref{spde_approximation}, we get $\phi_{\textrm{min}} = 0.76$ degrees and $\phi_{\textrm{max}} = 3.26$ degrees. With all the mesh nodes chosen as knots, i.e., $\mathcal{S}^{*(c)} = \mathcal{S}^*$ with $K=N^*$, we get $\phi_{\textrm{min}} = 0.25$ degrees and $\phi_{\textrm{max}} = 3.20$ degrees. If, additionally, $\mathcal{S} \subseteq \mathcal{S}^*$, we get $\phi_{\textrm{min}} = 0$ degree. Moreover, if we choose $K=25$ as in Figure~\ref{spde_approximation}, but set $\mathcal{S}^{*(c)} = \mathcal{S}^*$, we get $\phi_{\textrm{min}} = 1.46$ degrees and $\phi_{\textrm{max}} = 3.26$ degrees. The design shown in Figure~\ref{spde_approximation} thus allows obtaining a rather wide range of possible values for $\phi$, while keeping $K$ reasonably small.

We now describe the tail correlation coefficient of the scale mixture process \eqref{spatialscale_model}.

\begin{prop}\label{thm_tail_dependence_X}
For $\beta=0$, the tail-correlation coefficient of the $X(\cdot)$ process in \eqref{spatialscale_model} is
    \begin{eqnarray}
         \nonumber \chi_{X}(\bm{s}_1, \bm{s_2}) = \sum_{\substack{\{k: B_k(\bm{s}_1; \phi) > 0, \\ B_k(\bm{s}_2; \phi) > 0 \}}} \left[ B_k(\bm{s}_1; \phi) \bar{T}_{\gamma+1}\left( \sqrt{\gamma + 1} \ffrac{B_k(\bm{s}_1; \phi)^{1/\gamma} B_k(\bm{s}_2; \phi)^{-1/\gamma} - \rho(\bm{s}_1, \bm{s}_2)}{\sqrt{1 - \rho(\bm{s}_1, \bm{s}_2)^2}} \right) + \right. \\
        \nonumber \left. B_k(\bm{s}_2; \phi) \bar{T}_{\gamma+1}\left( \sqrt{\gamma + 1} \ffrac{B_k(\bm{s}_2; \phi)^{1/\gamma} B_k(\bm{s}_1; \phi)^{-1/\gamma} - \rho(\bm{s}_1, \bm{s}_2)}{\sqrt{1 - \rho(\bm{s}_1, \bm{s}_2)^2}} \right) \right],
    \end{eqnarray}
where $\bar{T}_{\rm df}(\cdot)$ denotes the Student's $t$ survival function with ${\rm df}$ degrees of freedom.   
\end{prop}

A formal proof is provided in the Supplementary Material. {Note that from straightforward copula theory, a similar result holds for the coefficient $\chi_{\tilde{X}}$ of the approximated process $\tilde{X}(\bm{s})$, where the only difference is that the true correlation function $\rho(\bm{s}_1, \bm{s}_2)$ has to be replaced with the GMRF-based approximate correlation structure.} Unlike $\chi_R$ in Proposition~\ref{thm_tail_dependence}, the coefficient $\chi_{X}$ in Proposition~\ref{thm_tail_dependence_X} does depend on $\gamma$, and also on the spatial correlation (\ref{cov_structure}) of the Gaussian process $Z(\cdot)$. Again, we have $\chi_X(\bm{s}, \bm{s})=1$ as expected, and if there is no basis function ``covering'' both $\bm{s}_1$ and $\bm{s}_2$, then $\chi_X(\bm{s}_1,\bm{s}_2)=0$. This implies that the $X(\cdot)$ process has local tail dependence, with the parameter $\phi$ controlling the range of the spatial tail dependence. However, unlike the $R(\cdot)$ process, which only allows for tail dependence with $\chi_R(\bm s_1,\bm s_2)>0$ or complete independence beyond a certain distance, the $X(\cdot)$ process captures sub-asymptotic tail dependence even when $\chi_X(\bm s_1,\bm s_2)=\chi_R(\bm s_1,\bm s_2)=0$ at large distances $\|\bm s_1-\bm s_2\|$, {due the presence of the Gaussian process $Z(\cdot)$ in its construction.} The dependence structure of the scale mixture process, $X(\cdot)$, is thus richer than that of the $R(\cdot)$ process itself, though the latter influences the former. Furthermore, while $X(\cdot)$ inherits a nonstationary dependence structure from $R(\cdot)$, an argument similar to the above discussion shows that $X(\cdot)$ is approximately stationary in a limiting sense as $K\to\infty$ and has tail-stationary margins for any finite $K$. In practice, the level of nonstationarity is dictated by the value of $K$ and the (relative) knot locations with respect to range parameter $\phi$.

To illustrate these properties, we consider the locations and mesh displayed in Figure~\ref{spde_approximation}, set $\phi=3$ or $6$ degrees, and consider three settings to construct the basis functions: (i) $K = 25$, $c = 1$; (ii) $K = N^*$, $c = 1$; and (iii) $K = 25$, $\mathcal{S}^{*(c)} \subset \mathcal{S}$ with $c=0.05$ (as shown in Figure~\ref{spde_approximation}). The other parameters are set to $\psi = 0.15\Delta_{\mathcal{S}}$, $r=0.9$, and $\nu=1$ for the correlation function $\rho(\bm{s}_1, \bm{s}_2)$ in \eqref{cov_structure}, and $\beta=0,\gamma = 2.5$ for the distribution of the random effects $R_k^*$ in \eqref{lowrank}. { The top panels of Figure~\ref{chi_X_illustration} show the tail-correlation coefficient $\chi_{R}(\bm{s}_1, \bm{s}_2)$ of the random scale process $R(\cdot)$ for each pair of locations, while the bottom panels show $\chi_{\tilde{X}}(\bm{s}_1, \bm{s}_2)$ (tail-correlation coefficient of the GMRF-based process $\tilde{X}(\cdot)$), plotted as a function of distance $d(\bm{s}_1, \bm{s}_2)$. When $\phi=3$, we can see that both $\chi_{R}$ and $\chi_{\tilde{X}}$ are identically zero for large distances as there is no basis function ``covering'' two locations $\bm{s}_1$ and $\bm{s}_2$ simultaneously. 
Alternatively, when $\phi=6$, for any pair of locations $\bm{s}_1$ and $\bm{s}_2$, there is at least one basis function $C_k$ that covers $\bm{s}_1$ and $\bm{s}_2$ simultaneously and hence both $\chi_{R}$ and $\chi_{\tilde{X}}$ are positive for all distances. Moreover, we can see that the first setting with $K=25$ and $c=1$ yields strong nonstationarity (seen through the high variability of $\chi_{R}$ and $\chi_{\tilde{X}}$ coefficients at a fixed distance), especially when $\phi=3$ degrees, while the other two settings are much closer to stationarity. We can also see that the degree of nonstationarity in the tail-dependence structure is weaker in the process $\tilde{X}(\cdot)$ than the process $R(\cdot)$.} 
In practice, it is, therefore, advisable to set $K=25$ and $c=0.05$, creating a nearly tail-stationary process, while keeping fast computations with a relatively small value of $K$.

\section{Bayesian inference}
\label{computation}

\subsection{Markov chain Monte Carlo sampling}
\label{mcmc_sampling}
Consider time-independent copies $\mathcal{Y} = \{Y_t(\bm{s}); \bm{s} \in \mathcal{S}, t=1,\ldots,T \}$ of the full model $Y(\bm s)$ in \eqref{full_model}. To each process $Y_t(\cdot)$, $t=1,\ldots,T$, corresponds independent and identically distributed replicates of the core process $X_t(\cdot)=R_t(\cdot)W_t(\cdot)$, with $R_t(\cdot)$ characterized by a collection of independent random effects $\bm{R}^*_t=[R_{1;t}^*,\ldots,R_{K;t}^*]'$, and $W_t(\cdot)$ constructed from independent GMRF replicates $\bm{\varepsilon}^*_{t}=[\varepsilon_{1;t},\ldots,\varepsilon_{N^*;t}]'$ and nugget terms $\bm{\eta}_t=[\eta_{1;t},\ldots,\eta_{N;t}]'$, which we need to infer together with the (time-invariant) hyperparameters.

We draw posterior inference about the hyperparameters and random effects 
using Markov chain Monte Carlo (MCMC) sampling, and we now give more details about our choice of priors and the proposed Bayesian inference procedure. Recall that in our application, the SPDE mesh is fixed to the one displayed in Figure~\ref{spde_approximation}, and we select different numbers of basis functions, namely $K = 3^2$, $4^2$, $5^2$, and $6^2$, while setting knots according to the maximum-minimum ordering with $\mathcal{S}^{*(c)} \subset \mathcal{S}$ with $c=0.05$ (see Section~\ref{r_construction}). Here we focus on the case where $\beta=0$ in \eqref{distr_r} with extremal range $\phi < \phi_{\max}$ in \eqref{basis_wendland} as this induces short-range AD and long-range AI. 


For the hyperparameters involved in the correlation function~\eqref{cov_structure} of $Z(\cdot)$, we choose the non-informative priors $\psi \sim \textrm{Uniform}(0, 2 \Delta_{\mathcal{S}})$ (correlation range) with $\Delta_{\mathcal{S}}$ the maximum spatial distance between pairs of locations, and $r \sim \textrm{Uniform}(0, 1)$ (nugget effect). 
The GMRF approximation is typically more accurate for smaller values of $\psi$ (weak dependence), and to get a good approximation for large values of $\psi$, as well, we define the SPDE mesh on an extended domain (recall Figure~\ref{spde_approximation}). For the hyperparameter $\gamma$ involved in \eqref{distr_r}, we choose the vague prior $\gamma \sim \textrm{Uniform}(0, 50)$. When $\beta=0$, $\gamma$ has a strong influence on the spatial extremal dependence (Theorem~\ref{thm_tail_dependence_X}). Small values of $\gamma$ induce a large value of $\chi_X$ (strong tail dependence), while large $\gamma$ leads to $\chi_X\approx 0$ (weak tail dependence) at any distance. 
We find that it is very challenging to efficiently update the Wendland basis range parameter $\phi$ (extremal range), which might be due to the high posterior correlation with other model parameters. Therefore, we fix $\phi$ to some predefined values when running the MCMC algorithm and select the best value afterward using information criteria. Specifically, we set it to $\phi = 0.75\phi_{\min} + 0.25 \phi_{\max}$, $\phi = 0.5 \phi_{\min} + 0.5 \phi_{\max}$, and $\phi = 0.25 \phi_{\min} + 0.75 \phi_{\max}$, for which we run separate MCMC chains, and then choose both $K$ and $\phi$ using the deviance information criterion (DIC). We model the mean process $\mu(\cdot)$ as $\mu(\bm{s}) = D(\bm{s})'\bm{\theta} + \varepsilon_{\mu}(\bm{s})$, $\bm s\in\mathcal{S}$, where the covariates are $D(\bm{s}) = [1, \textrm{lon}(\bm{s}), \textrm{lat}(\bm{s}), \textrm{elev}(\bm{s})]'$ for each location $\bm{s}$ and specify $\varepsilon_{\mu}(\bm{s}) \overset{\textrm{IID}}{\sim} \textrm{Normal}(0, \tau_{\mu}^{-1})$ with precision parameter $\tau_{\mu}>0$. Hyperpriors are chosen as $\bm{\theta} \sim \textrm{Normal}(\bm{0}, 100^2 \bm{I}_4)$ and $\tau_{\mu} \sim \textrm{Gamma}(0.1, 0.1)$. Because our main focus is on inference and not on spatial prediction, a spatially independent prior for $\mu(\cdot)$ is reasonable. We also specify a non-informative prior for the spatially-constant precision parameter $\tau$ in the full model~\eqref{full_model}, namely $\tau \sim \textrm{Gamma}(0.1, 0.1)$. { We choose the above-mentioned priors for both the simulation study and the real data application.}

The MCMC steps are quite standard: we update the hyperparameters $\mu(\bm{s}_i)$, $\bm{\theta}$, $\sigma^2_{\mu}$, $\sigma^2$ using Gibbs sampling, and $\psi$, $\gamma$, and $r$ using an adaptive Metropolis--Hastings algorithm. For the latent random effects, we update $\bm{\varepsilon}_t^*$, $t=1, \ldots, T$, using Gibbs sampling, and $\bm{R}^*_{t}$, $t=1, \ldots,T$, using an adaptive Metropolis-adjusted Langevin algorithm (MALA). Because of the independence assumption across time in (\ref{full_model}), we update  $\bm{\varepsilon}_t^*$ and $\bm{R}^*_{t}$ (separately for each $k$th vector component) in parallel. The sparse SPDE approximation of $Z(\cdot)$ allows fast updating of the vectors $\bm{\varepsilon}_t^*$. Furthermore, the low-rank construction of $R(\cdot)$, with compactly-supported basis functions, allows fast updating of $\bm{R}^*_{t}$. Finally, we impute the censored observations below the thresholds jointly using Gibbs sampling, thanks to the nugget term $\bm{\eta}$ in \eqref{z_construction_eq} allowing fast univariate updates. Full conditional distributions are detailed in the Supplementary Material.

We implement the MCMC algorithm in \texttt{R} (\url{http://www.r-project.org}). { For the data application, we generate 20 parallel chains with $100\,000$ iterations each, and discard the first $50\,000$ iterations as burn-in period. Subsequently, we thin each Markov chain by keeping one out of five consecutive samples, and thus, we finally obtain $10\,000$ posterior samples from each chain and overall $200\,000$ samples to draw posterior inference.} In our simulation study, we instead generate a single chain, where we halve both the burn-in and post-burn-in periods to $25\,000$ iterations for computational reasons, but we keep the same thinning factor (thus yielding $5\,000$ posterior samples). Convergence and mixing are monitored through trace plots.  

\subsection{Simulation study}
\label{simulation}

\begin{figure}[t!]
\centering
	\adjincludegraphics[width = \linewidth, trim = {{.0\width} {.0\width} {.0\width} {.0\width}}, clip]{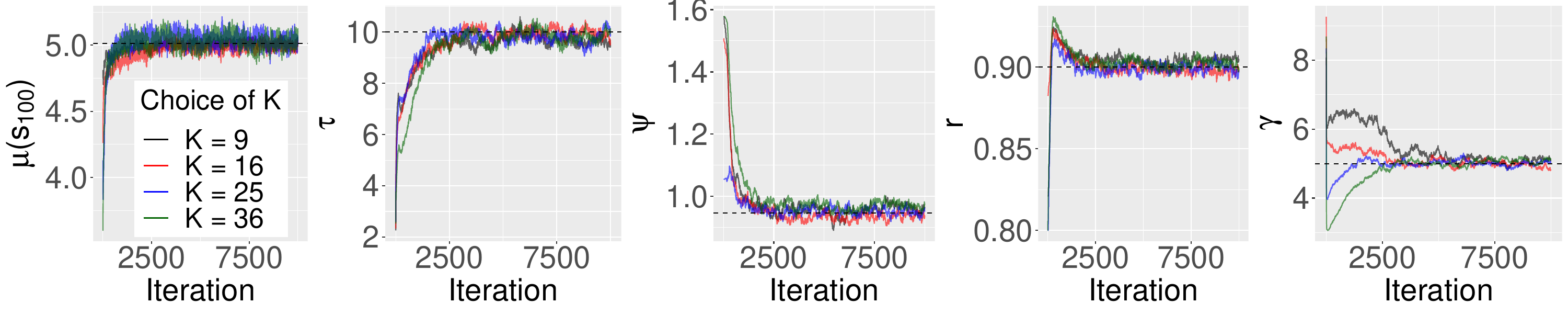}
	\caption{Trace plots (thinned MCMC chains) of some of the model parameters. The corresponding true parameter choices are shown using the dashed horizontal lines.}
	\label{trace_plots}
\end{figure}

In this section, we demonstrate the performance of the proposed model and our Bayesian inference approach in terms of convergence diagnostics for the MCMC chains and coverage probabilities for the true pairwise $\chi$-measure. In our simulation study design, we mimic our real data application, and choose the same spatial domain and number of time replicates; precisely, we simulate datasets at $195$ grid cells over Bangladesh at a resolution of $0.25^\circ \times 0.25^\circ$ with $2440$ independent time replicates from the full model \eqref{full_model}, setting $K=3^2$, $4^2$, $5^2$ and $6^2$. True parameters are set according to $\mu(\bm{s}) = 5 + 0.25\,\textrm{lon}(\bm{s})^2 + 0.25\,\textrm{lat}(\bm{s})^2 + 0.25\,\textrm{elev}(\bm{s})^2$ (thus depending quadratically on the longitude/latitude coordinates and the elevation of each location $\bm s$), $\tau = 10$, $\psi = 0.15\Delta_{\mathcal{S}} = 0.946$, $r = 0.9$, and $\gamma = 5$ (i.e., with marginal tail index $1/5$). Moreover, we fix the extremal range to $\phi = \frac{3}{4}\phi_{\min} + \frac{1}{4} \phi_{\max}$, which resembles the results in our data application. At each spatial location, we censor the observations below the $95$th data percentile and impute the censored observations. Some trace plots are presented in Figure~\ref{trace_plots}. 
We display the $10\,000$ MCMC samples obtained after thinning, where the first $5\,000$ represent the burn-in period. Even though the starting values are quite far from the true parameter values, all trace plots converge relatively fast. Despite a large number of latent variables requiring Metropolis--Hastings updates and left-censoring, the MCMC chains mix reasonably well. Estimating the shape parameter $\gamma$ in (\ref{distr_r}) appears to be the most challenging task and a moderately large burn-in period is thus required. {We repeat the experiment 30 times under each setting of $K=3^2$, $4^2$, $5^2$, and $6^2$ and calculate the effective sample size (ESS) for the same set of parameters presented in Figure~\ref{trace_plots}. In Table~\ref{ess_simulation}, we report the average and standard deviation of ESS values across the 30 simulated datasets, for a set of model parameters and under different choices of $K$. 
\begin{table}[t!]
\caption{{Average and standard deviation (within brackets) of effective sample sizes across 30 simulated datasets, for a set of model parameters and under different choices of $K$.}}\vspace{5pt}
\centering
\begin{tabular}{c | ccccc}
\hline
Choice of $K$ & $\mu(\bm{s}_{100})$ & $\tau$ & $\psi$ & $r$ & $\gamma$ \\ 
\hline
$K=3^2$ & 104.31 (42.09) & 16.19 (7.77) & 19.13 (5.56) & 37.25 \hphantom{0}(9.78) & 19.11 (8.43) \\ 
$K=4^2$ & \hphantom{0}72.62 (29.67) & 19.29 (9.87) & 20.91 (5.12) & 36.91 (10.63) & 10.41 (4.10) \\ 
$K=5^2$ & 103.88 (43.94) & 15.09 (6.80) & 20.93 (6.31) & 32.67 \hphantom{0}(7.57) & \hphantom{0}6.99 (2.81) \\ 
$K=6^2$ & \hphantom{0}99.94 (40.58) & 15.57 (5.74) & 21.16 (6.10) & 34.96 \hphantom{0}(7.87) & \hphantom{0}7.37 (4.13) \\ 
\hline
\end{tabular}
\label{ess_simulation}
\end{table}
The ESS for $\gamma$ is very low, which is consistent with our findings from Figure~\ref{trace_plots}. Running a (single) longer MCMC chain or (multiple) parallel MCMC chains would be required for more accurate inference. While we do not implement such improvements in this simulation study due to limited computational resources, we run 20 parallel chains in our real data application in Section~\ref{application}. From the single Markov chain in this simulation study, we then compute the posterior mean and standard deviation of the model parameters and report in Table~\ref{posmeansd} their average values across the $30$ simulated datasets. For each parameter presented in Table~\ref{posmeansd}, the average of the posterior means is close to the true parameter value and the average of the posterior standard deviations is fairly small.}

\begin{table}[t!]
\caption{{Posterior means and standard deviations (in brackets) for some model parameters, averaged across the $30$ simulated datasets.} The true values of the corresponding parameters were set to $\mu(\bm{s}_{100}) = 5.014$, $\tau = 10$, $\psi = 0.946$, $r = 0.9$, and $\gamma = 5$.}\vspace{5pt}
\centering
\begin{tabular}{c | ccccc}
\hline
Choice of $K$ & $\mu(\bm{s}_{100})$ & $\tau$ & $\psi$ & $r$ & $\gamma$ \\ 
\hline
$K=3^2$ & 5.008 (0.021) & 9.678 (0.223) & 0.954 (0.013) & 0.901 (0.002) & 5.075 (0.095) \\ 
$K=4^2$ & 5.015 (0.029) & 9.745 (0.218) & 0.953 (0.013) & 0.901 (0.002) & 5.029 (0.097) \\ 
$K=5^2$ & 5.034 (0.043) & 9.764 (0.218) & 0.951 (0.013) & 0.901 (0.002) & 4.979 (0.100) \\ 
$K=6^2$ & 5.033 (0.043) & 9.737 (0.212) & 0.953 (0.013) & 0.901 (0.002) & 4.993 (0.087) \\
\hline
\end{tabular}
\label{posmeansd}
\end{table}


\begin{figure}[h]
\centering
	\adjincludegraphics[width = 0.8\linewidth, trim = {{.0\width} {.0\width} {.0\width} {.0\width}}, clip]{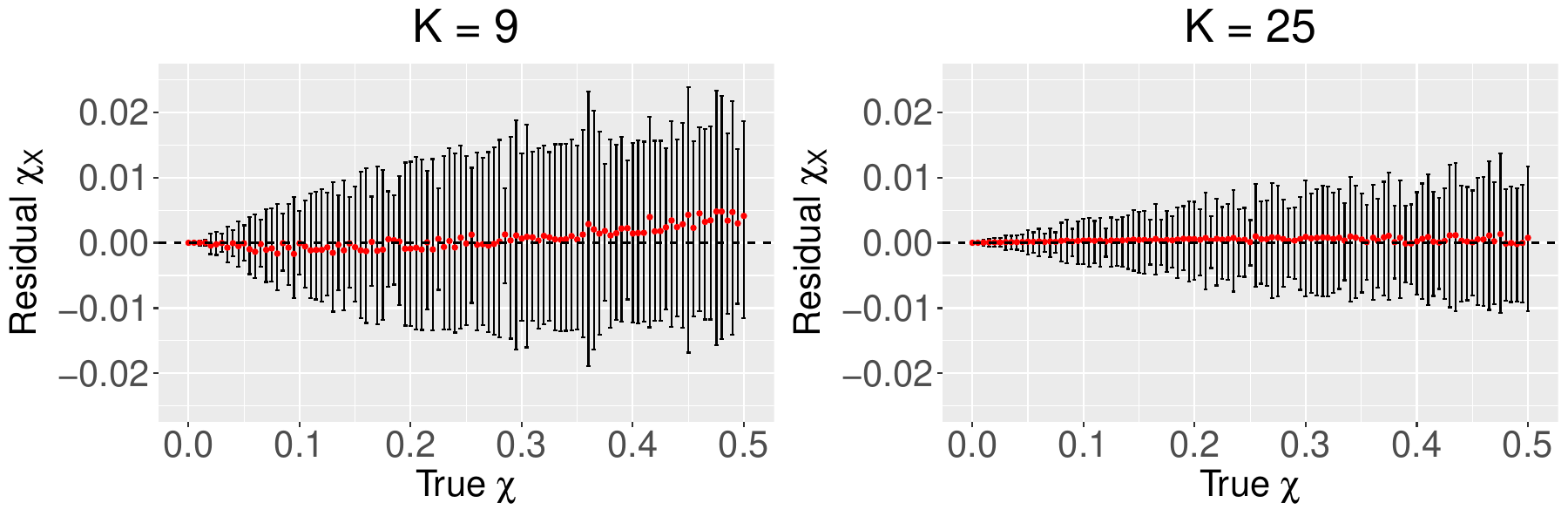}
	\vspace{-4mm}
	\caption{Residual $\chi_X$ ({dots}), i.e., the difference between the posterior mean of $\chi_X$ and the true $\chi_X$, with its corresponding $95\%$ posterior credible intervals (vertical bars) {averaged across the 30 simulated datasets}, for some pairs of sites with small to high values {of the true $\chi_X$}, for $K=9$ (left) and $K=25$ (right).}
	\label{chi_credible}
\end{figure}

For a representing subset of pairs of sites, we then compare the true pairwise $\chi_X$-measure using \eqref{thm_tail_dependence_X} with its model-based counterpart, estimated as the posterior mean using posterior samples, and we also compute the corresponding $95\%$ credible intervals{, and average them across the 30 simulated datasets. Here, we consider a grid of values of true $\chi$ as $\{0, 0.005, 0.01,\ldots, 0.5\}$ and for each $\chi$, we pick the pair of sites whose theoretical $\chi$-measure, calculated based on the true parameter choices under the simulation design, is nearest to that $\chi$ value. Based on our fitted model, we then also estimate the corresponding $\chi$-measure for that pair of sites (along with a 95\% credible interval),} and Figure~\ref{chi_credible} shows the estimation error (i.e., the difference between true and estimated $\chi_X$-measures, referred to here as the ``residual $\chi_X$'') and its posterior uncertainty for $K=3^2$ and $K=5^2$, plotted as a function of the true $\chi_X$-measure. All credible intervals {(averaged across 30 datasets)} contain zero and they are wider for larger $\chi_X$, as expected, and for smaller $K$ (fewer basis functions). The {(averaged)} credible intervals for the highest true $\chi_X$ values span approximately between $(-0.015, 0.02)$ for $K=9$ and $(-0.01, 0.01)$ for $K=25$, so the estimation error is nevertheless quite small.

\section{Data application: rainfall extremes in Bangladesh}
\label{application}

\subsection{The precipitation dataset and exploratory analysis}
\label{data}

We analyze daily precipitation data obtained from the project Tropical Rainfall Measuring Mission (TRMM, Version 7, \url{https://gpm.nasa.gov/data-access/downloads/trmm}) available over the period March 2000 through December 2019 at a spatial resolution of $0.25^\circ \times 0.25^\circ$; considering the whole Bangladesh, we have data available at $195$ grid cells. In order to study heavy rainfall that affects monsoon crops, we consider data only for the months of June to September and thus, we finally obtain $2440$ temporal replicates. 

\begin{figure}[!t]
\centering
\adjincludegraphics[height = 0.33\linewidth , trim = {{.0\width} {.0\width} {.40\width} {.0\width}}, clip]{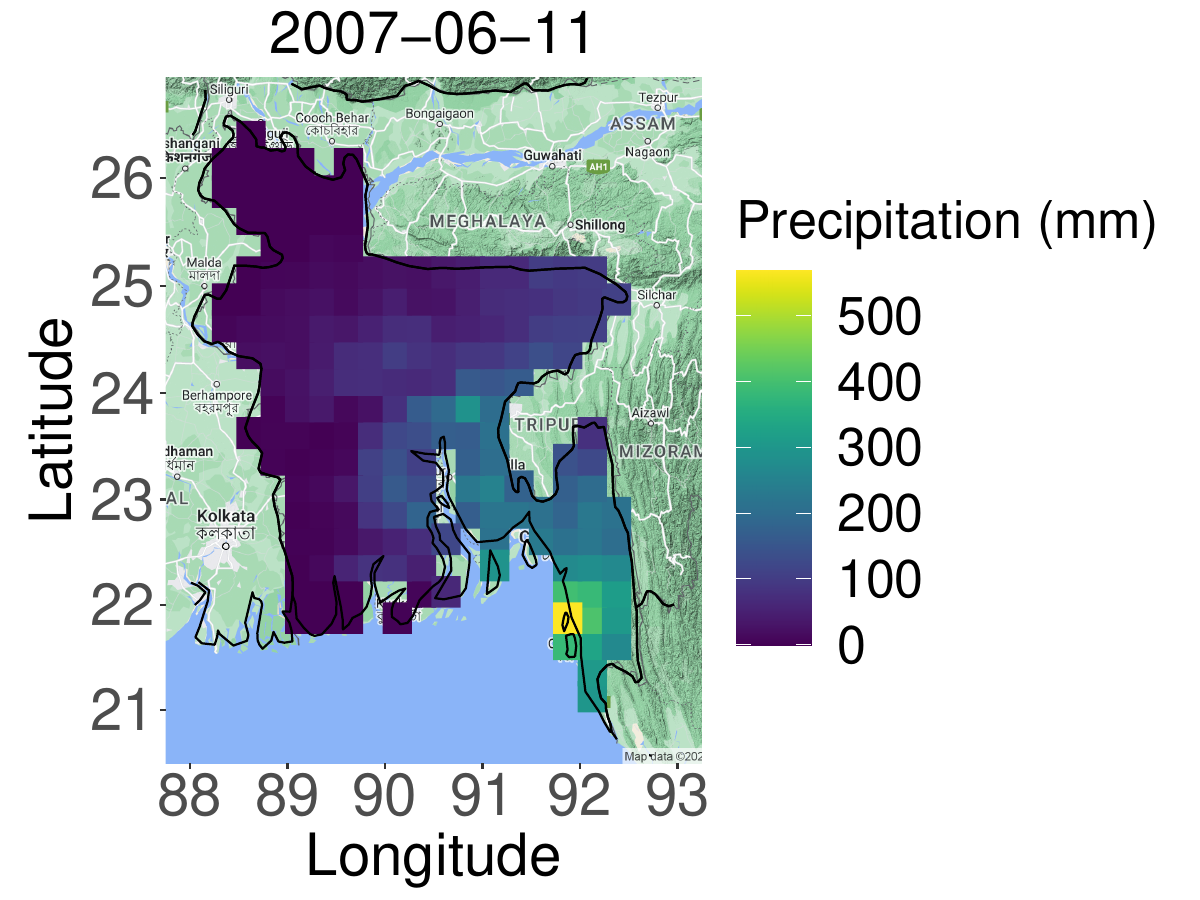}
\adjincludegraphics[height = 0.33\linewidth, trim = {{.0\width} {.0\width} 0 {.0\width}}, clip]{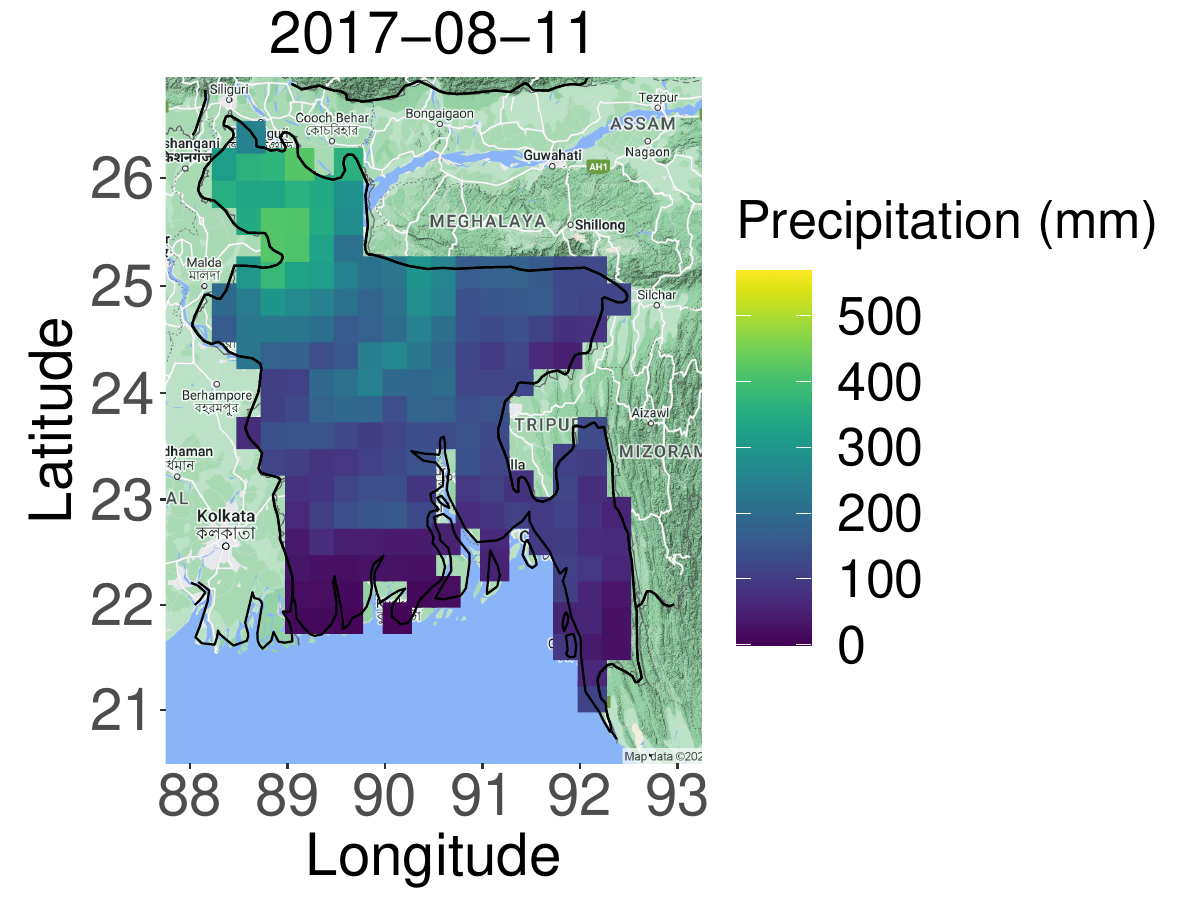}
\vspace{-2mm}
\caption{Spatial maps of daily precipitation for the day experiencing the highest precipitation amount at a single grid cell (left) and the day with the highest spatially averaged precipitation (right). Both panels are plotted on the same color scale.}
\label{high_two_cases}
\end{figure}

Figure~\ref{high_two_cases} displays spatial maps of daily precipitation, for the day experiencing the highest precipitation amount at a single grid cell (left panel) and the day with the highest spatially averaged precipitation (right panel). From the left panel, we see that the highest extreme events tend to be spatially localized. From the right panel, we see moderately high precipitation values that cover a much wider region, which decrease gradually with distance from the storm center. This suggests our spatial model, which can capture both short-range AD and long-range AI through latent local shocks, may be a suitable choice.  

Because our full model~\eqref{full_model} has a spatially-constant precision parameter $\tau$, we first center and scale the data by subtracting the median of the positive observations at each site followed by dividing the residuals by the interquartile range (IQR) of the positive observations. As our main goal is in estimating the shape of the tail and the extremal dependence, this simple preprocessing step is justified. The final results are, however, presented on their original scale. Note that we fit the model to data exceeding $95$th marginal percentiles through a censored inference algorithm; therefore, the centering and rescaling do not influence the tail directly. 

A detailed exploratory analysis is provided in the Supplementary Material, so we here only summarize our main findings. Simple diagnostics and tests suggest that threshold exceedances do not exhibit any significant lag-$1$ autocorrelation. Moreover, our analysis suggests that the assumption of year-by-year stationarity over the period under study is reasonable. 
Thus, we safely ignore the modeling of temporal dependence and trends in our analysis hereafter, and simply treat time replicates as independent and identically distributed spatial realizations. Further, we compare the marginal fits from a generalized Pareto distribution (GPD) with that of the marginal distribution of the Gaussian scale mixture \eqref{full_model}; if the SHOT model in \eqref{full_model} fits the margins equally well or better than a GPD, we can then avoid a copula-based approach. Although the SHOT model has a slightly larger RMSE than the GPD in high quantiles, discrepancies are quite small. Hence, we consider a unified approach for modeling margins and dependence simultaneously using the SHOT model, as this facilitates both computations and interpretation. We finally also explore the data's extremal dependence structure and compute the empirical $\chi$-measure between all the pairs of grid cells. At large distances, empirical estimates show evidence of extremal independence, while extremal dependence is substantially stronger at small distances; this indicates that we need a model that flexibly captures AD and AI, potentially changing as a function of the distance between sites.

\subsection{Model comparison}
\label{model_comparison}

In this section, we compare our proposed SHOT model \eqref{full_model} to a GMRF and the single-scale mixture HOT model constructed from \eqref{hot_w}, all fitted to spatial threshold exceedances using the same censored inference algorithm. {Prior specification for the model parameters and hyperparameters are as described in Section \ref{mcmc_sampling}.} To be consistent across models, we also use a GMRF for constructing the HOT process and consider the same mesh in all three models. For the proposed model, we consider $K = 3^2$, $4^2$, $5^2$, and $6^2$ basis functions and extremal range $\phi = {3\over4}\phi_{\min} + {1\over4} \phi_{\max}$, ${1\over2}\phi_{\min} + {1\over2} \phi_{\max}$, and ${1\over4}\phi_{\min} + {3\over4} \phi_{\max}$. First, we compare the different models based on (scaled) DIC; a model with a smaller value of DIC is preferred. For readability, we divide the final DIC values by $475\,800$, the total number of spatiotemporal observations. The DIC values for the GMRF and HOT models are $-6.560$ and $-7.644$, respectively. The DIC values for the SHOT model with respect to $K$ and $\phi$ are reported in Table~\ref{dic_table}. The best SHOT model is obtained with $K=25$ and $\phi = {3\over4}\phi_{\min} + {1\over4} \phi_{\max}$, and is equal to $-9.276$, a substantial improvement with respect to the GMRF and the HOT models. We thus use the SHOT model with these values of $K$ and $\phi$ to draw inferences.

\begin{table}[t!]
\caption{Deviance information criterion (DIC) of the proposed SHOT model for different choices of $K$ and $\phi$. A smaller value of DIC is preferred.}\label{dic_table}\vspace{5pt}
\centering
\begin{tabular}{c | ccc}
  \hline
Choice of $K$ & $\phi = \frac{3}{4}\phi_{\min} + \frac{1}{4} \phi_{\max}$ & $\phi = \frac{1}{2}\phi_{\min} + \frac{1}{2} \phi_{\max}$ & $\phi = \frac{1}{4}\phi_{\min} + \frac{3}{4} \phi_{\max}$ \\ 
  \hline
$K=3^2$ & -7.583 & -7.394 & -7.875 \\ 
$K=4^2$ & -8.098 & -8.502 & -8.014 \\ 
$K=5^2$ & \textbf{-9.276} & -8.746 & -8.552 \\ 
$K=6^2$ & -8.710 & -8.759 & -8.168 \\
  \hline
\end{tabular}
\end{table}


{The computation times for the GMRF, HOT model, and SHOT model with $K=3^2$, $4^2$, $5^2$, and $6^2$ (averaged across different choices of $\phi$) are $513.37$, $534.36$, $560.11$, $575.85$, $595.21$, and $618.55$ minutes, respectively (with $20$ parallel MCMC chains run simultaneously, but one-at-a-time for each model, on a workstation with AMD Ryzen 9 processor and 64 GB RAM)}. Although our proposed SHOT model is substantially more complicated than the GMRF and has many additional latent variables, which are updated using the adaptive Metropolis-adjusted Langevin algorithm, the computation time for the SHOT model with $K=5^2$ is only $1.24$ times that of the GMRF. Our implementation is thus relatively optimized. 

\begin{figure}[t!]
\centering
	\adjincludegraphics[width = \linewidth, trim = {{.0\width} {.0\width} {.0\width} {.0\width}}, clip]{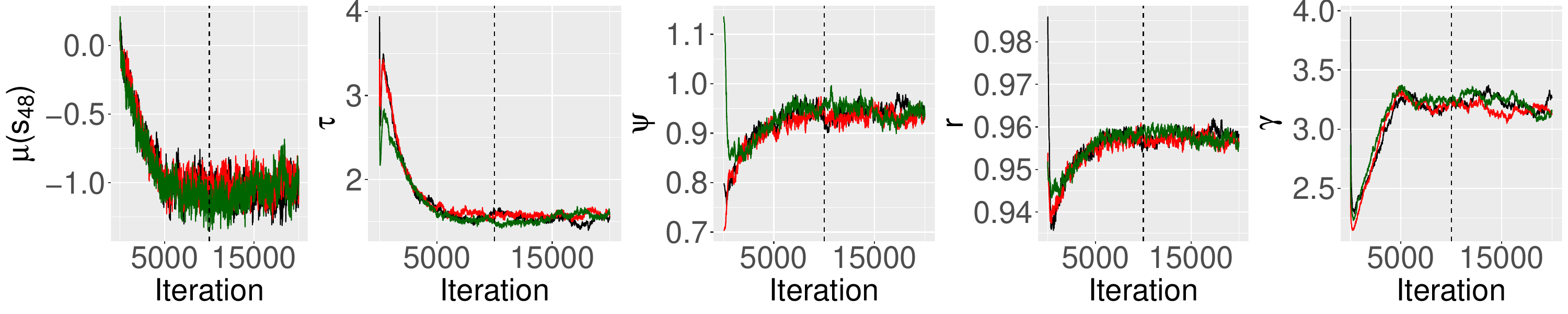}
	\caption{Trace plots ({three different thinned MCMC chains with different starting values}) of some parameters for the best SHOT model with $K=5^2$ and $\phi = {3\over4}\phi_{\min} + {1\over4} \phi_{\max}$. The burn-in period (dashed vertical line) is set to $10\,000$ iterations (after thinning).}
	\label{trace_plots_real}
\end{figure}

Trace plots for some model parameters of the best SHOT model, i.e., when $K=5^2$ and $\phi = \frac{3}{4}\phi_{\min} + \frac{1}{4} \phi_{\max}$, are presented in Figure~\ref{trace_plots_real}. Overall, the MCMC chains mix well and converge after a relatively short number of iterations. Similar to our simulation study, the convergence of the MCMC chains for the parameter $\gamma$ appears to be the most challenging, and a moderately large burn-in period is thus required. 
Posterior means, standard deviations, and $2.5$th and $97.5$th posterior percentiles for the hyperparameters $\tau$, $\psi$, $r$, $\gamma$, and $\mu(\cdot)$ at the pixel $\bm{s}_{48}$ with centroid (22.875$^\circ$N, 91.875$^\circ$E), are reported in Table~\ref{pos_summaries}. {In particular, $r$ has posterior mean $0.957$ (small nugget), and the estimated $\gamma$ is $3.165$, indicating a marginal tail index $0.32$, which corresponds to moderately heavy tails as expected for precipitation data.}


\begin{table}[t!]
\caption{{Posterior means, standard deviations, and $2.5$th and $97.5$th posterior percentiles for some parameters of the SHOT model with $K=5^2$ and $\phi = \frac{3}{4}\phi_{\min} + \frac{1}{4} \phi_{\max}$.}}\vspace{5pt}
\centering
\begin{tabular}{lccccc}
  \hline
Parameter & $\mu(\bm{s}_{48})$ & $\tau$ & $\psi$ & $r$ & $\gamma$ \\ 
  \hline
Posterior mean & -1.021 & 1.568 & 0.939 & 0.957 & 3.165 \\ 
  Posterior standard deviation & 0.088 & 0.048 & 0.014 & 0.001 & 0.069 \\ 
  $2.5$th posterior percentile & -1.191 & 1.477 & 0.912 & 0.955 & {3.039} \\ 
  $97.5$th posterior percentile & -0.846 & 1.667 & 0.966 & 0.960 & {3.305} \\ 
   \hline
\end{tabular}
\label{pos_summaries}
\end{table}

\begin{figure}[t!]
\centering
	\includegraphics[height = 0.33\linewidth]{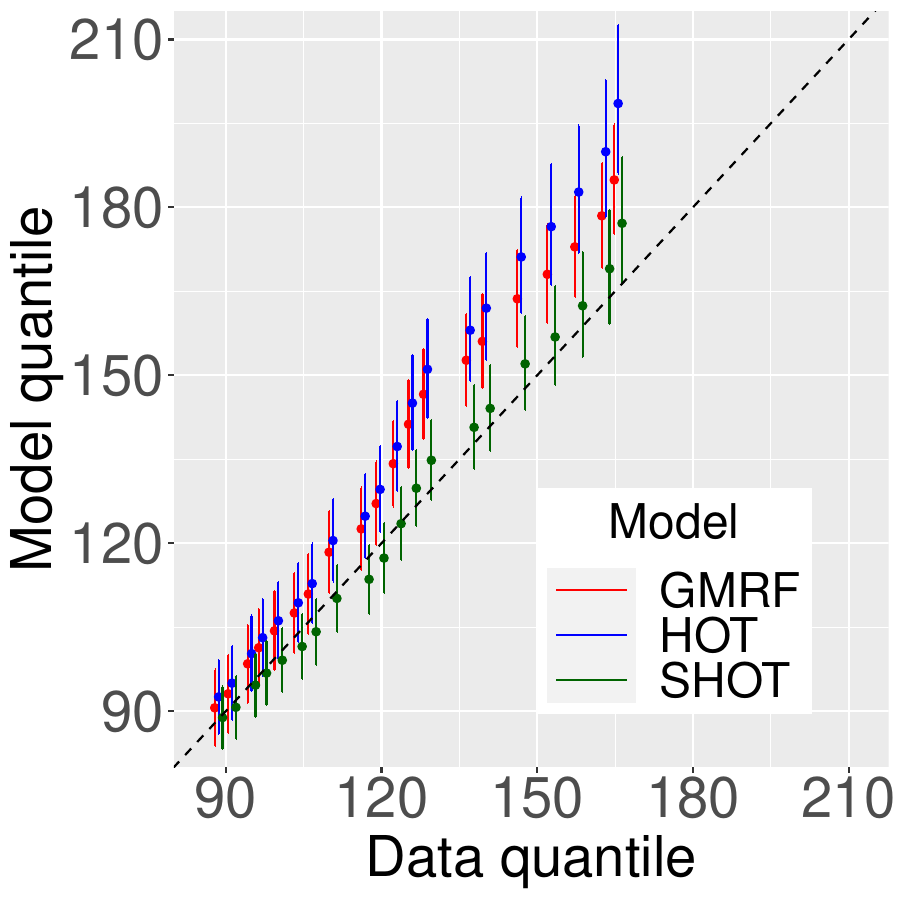}
	\includegraphics[height = 0.33\linewidth]{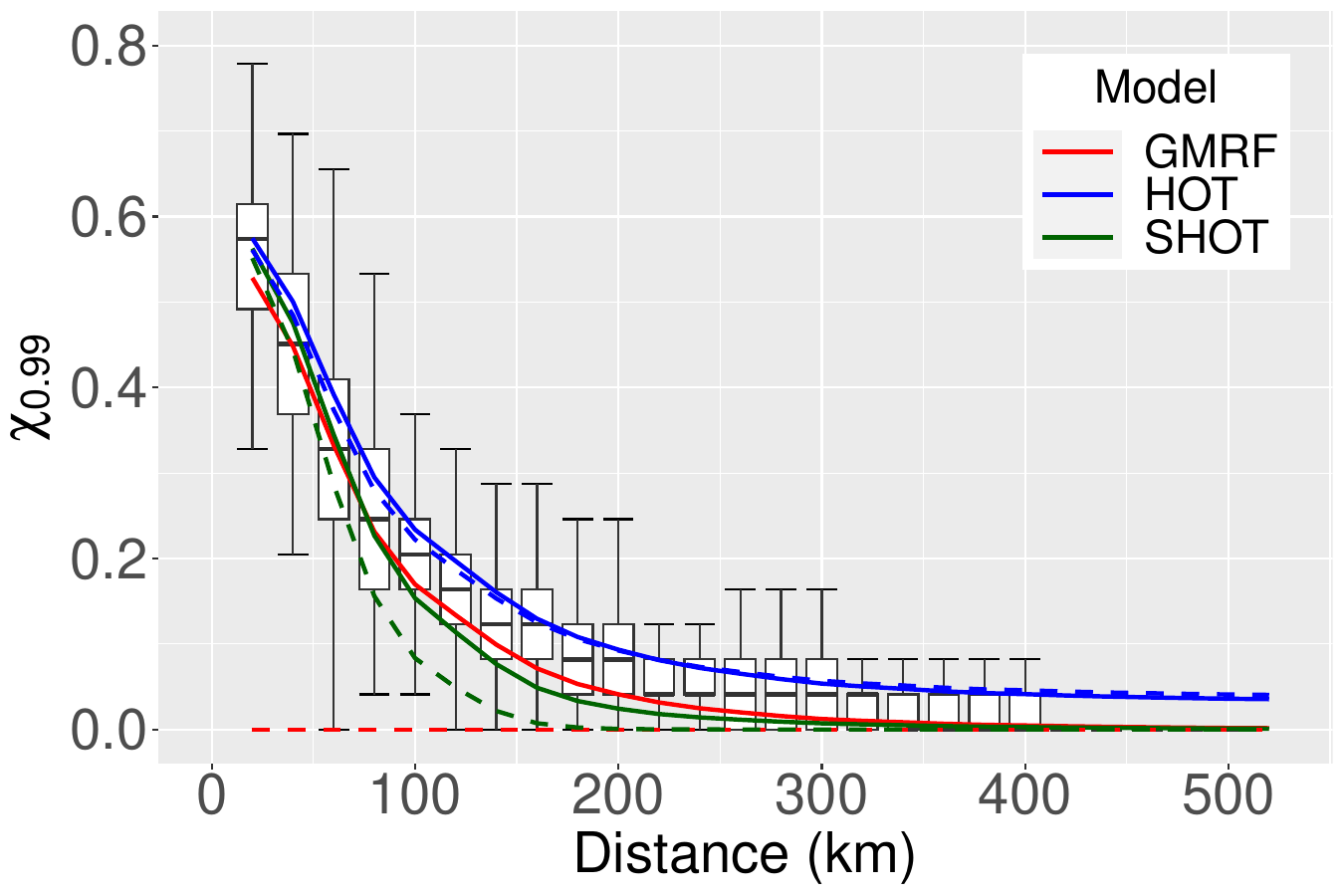}
	\caption{\emph{Left:} QQ-plot (marginal model-based quantiles vs data quantiles) for levels between $0.95$ and $0.995$, at the pixel with coordinate (22.875$^\circ$N, 91.875$^\circ$E). The dots represent the corresponding posterior medians and the bars represent the corresponding 95\% credible intervals. \emph{Right:} Boxplots of empirical $\chi_u$ for $u=0.99$, for low through high distances $(20\text{km}, 40\text{km},\ldots,$ and bandwidth of 20km). The overlapped solid lines represent the fitted $\chi_{0.99}$ curve based on different models and the dashed lines represent the theoretical $\chi$ as a function of distance.}
	\label{quantile_chi}
\end{figure}

To assess the marginal model goodness-of-fit, we then compare high marginal data quantiles to the corresponding model quantiles. The left panel of Figure~\ref{quantile_chi} shows a quantile-quantile (QQ) plot for the GMRF, HOT, and SHOT models (with quantiles estimated by simulation) for the pixel with coordinates (22.875$^\circ$N, 91.875$^\circ$E), focusing on the upper tail (for quantile levels between $0.95$ and $0.995$). 
For quantile levels close to $0.95$, the model-based quantiles (estimated through their posterior medians) are close to the data quantiles for all three models. However, at higher quantile levels, the model-based quantiles are only accurate for the SHOT model while they are significantly overestimated for the GMRF and HOT models. 

To assess the fit of the extremal dependence structure, we then compute the tail correlation coefficient $\chi_u$ with $u=0.99$ as in \eqref{eq:chi} for all pairs of sites, as well as its limit $\chi_X$, both empirically and based on the three fitted models (by simulation). The right panel of Figure~\ref{quantile_chi} shows the results. Empirical $\chi$ estimates are displayed as boxplots for distance classes ranging from $20$km to $500$km, while model-based $\chi_{0.99}$ estimates are shown as solid curves. {Specifically, for each distance in $\{20\text{km}, 40\text{km},\ldots,500\text{km}\}$, we consider a bandwidth of $20$km, i.e., with distance classes defined as the intervals $(10\text{km}, 30\text{km})$ up to $(490\text{km}, 510\text{km})$; we then display boxplots of the empirical $\chi$ estimates for all pairs of locations with geodesic distance falling into such intervals.} All models perform quite well overall, especially at short distances. However, while the GMRF and the SHOT models slightly underestimate the empirical $\chi_{0.99}$ estimates at moderate distances, the HOT model has the opposite behavior and consistently overestimates it. Moreover, both the GMRF and SHOT models realistically capture full independence at long distances, while the HOT model is unable to capture it due to its construction in terms of a single spatially-constant random scale variable. Finally, the limiting $\chi$-measure is non-trivial for the HOT and SHOT models at short-to-moderate distances, while it is zero at any distance for the GMRF. Thus, the SHOT model has the most appealing dependence properties.

Overall, from the DIC values in Table~\ref{dic_table} and the marginal and dependence diagnostics in Figure~\ref{quantile_chi}, the SHOT model is the best and most realistic one compared to its competitors.

\subsection{Marginal and spatially aggregated precipitation risk estimates}
\label{inference}
\begin{figure}[t!]
\centering
	\adjincludegraphics[width = 1.1\linewidth, trim = {{.07\width} {.42\width} {.05\width} {.41\width}}, clip]{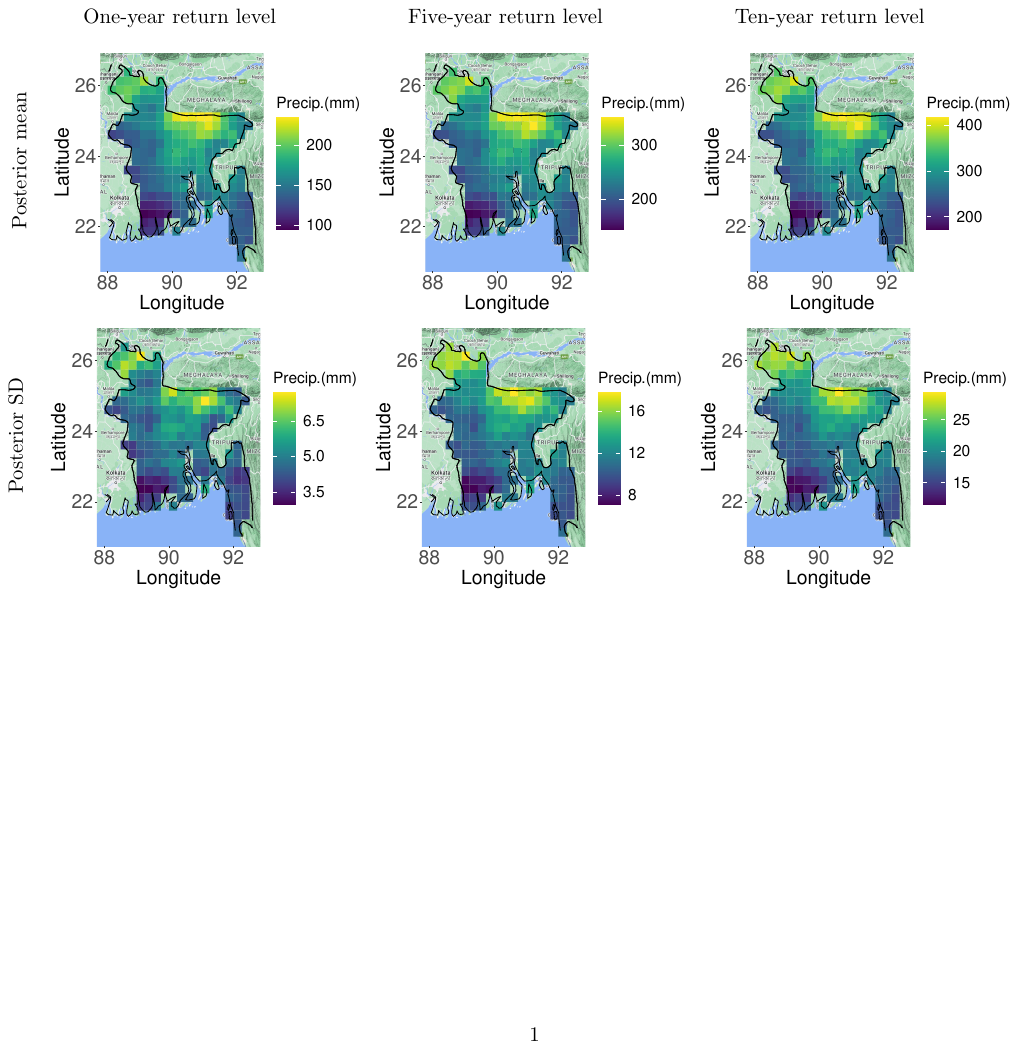}
	\caption{Spatial maps of posterior means (top) and standard deviations (bottom) of $1$-year (left), $5$-year (middle), and $10$-year (right) return levels, computed from the best SHOT model.}\label{return_levels}
\end{figure}

{We finally use our best-fitting SHOT model to assess marginal and spatially aggregated risk, respectively defined as return levels for precipitation observed at each individual spatial location, as well as for precipitation averaged over specific sub-regions.} 
The $m$-year return level here corresponds to the $\{1-1/(122m)\}$-quantile given that there are only $122$ days per year during the period from June to September. Figure~\ref{return_levels} displays spatial maps of marginal $m$-year precipitation return levels for $m=1,5,10$ at each site. The posterior means of $1$-year return levels vary between $94.71$mm and $233.40$mm, with the smallest values observed near the southwestern region and the highest values observed near the Himalayan foothills of the northeastern region. The posterior means of $5$-year return levels vary between $144.35$mm and $348.96$mm, while that for the $10$-year return levels vary between $172.14$mm and $415.01$mm, with similar spatial patterns in all cases. The posterior standard deviations of return levels are quite moderate overall, thanks to our spatial model which borrows strength across neighboring locations.
We then further estimate return levels of spatially aggregated precipitation for six sub-regions of Bangladesh, whose delineation was chosen according to the definition used in  \cite{mannan2007climatic} who analyzed heavy rainfall in Bangladesh. Table~\ref{return_level_regional} reports the posterior mean and standard deviation for the $m$-year return level, with $m=1,5,10$, {along with empirical $1-\textrm{year}$ return levels}, of the spatial average for each region. {For empirical $5$-year and $10$-year return levels, only 4 and 2 observations exceed the corresponding values and the estimates involve very high uncertainty; we thus do not report them here. The empirical $1$-year return levels, although being also quite uncertain, are in reasonably good agreement with the model-based $1$-year return levels, confirming a satisfactory goodness-of-fit.} As expected, the posterior mean is the highest for the North-East region while it is the smallest for the South-West region (for all $m=1,5,10$), and this pattern is consistent with the spatial maps presented in Figure~\ref{return_levels}. The posterior standard deviations are again relatively small with respect to the scale of return level estimates. Such results could be exploited for assessing flood risk and, in principle, used by local authorities for regional flood mitigation and planning. 

\begin{table}[t!]
\caption{Posterior means and standard deviations (within brackets) of $1$-year, $5$-year, and $10$-year return levels (RL; in mm) of the spatially averaged precipitation within six geographical regions of Bangladesh, computed by simulation from the best SHOT model. { We also report the empirical $1$-year return levels in the second column.}}\label{return_level_regional}\vspace{5pt}
\centering
\begin{tabular}{l| c | ccc}
  \hline
Region & $1$-year RL & $1$-year RL & $5$-year RL & $10$-year RL \\ 
\hline
 & Empirical & & Model-based & \\
  \hline
North-West & 109.031 & 109.255 (3.396) & 160.456 (7.141) & 186.391 (10.993) \\ 
  North-East & 140.646 & 140.084 (3.928) & 200.266 (8.462) & 230.796 (12.951) \\ 
  West-Central & \hphantom{0}95.591 & 110.169 (3.560) & 165.355 (7.591) & 193.296 (11.683) \\ 
  East-Central & 110.700 & 116.081 (3.396) & 166.886 (7.022) & 192.281 (10.732) \\ 
  South-West & \hphantom{0}98.091 & \hphantom{0}88.749 (2.850) & 132.596 (6.068) & 154.756 \hphantom{0}(9.283) \\ 
  South-East & 118.496 & 103.207 (2.894) & 148.212 (6.370) & 171.636 \hphantom{0}(9.981) \\
   \hline
\end{tabular}
\end{table}

\section{Discussion}
\label{conclusion}

We have developed a novel Bayesian hierarchical model for spatial extremes---called the SHOT model---which can capture short-range asymptotic dependence (AD), mid-range asymptotic independence (AI), and long-range exact independence. This model is the first-of-its-kind that can smoothly bridge AD and AI as a function of spatial distance, and that also has a global unconditional stochastic representation, while allowing for feasible and relatively fast fully-Bayesian inference in high spatial dimensions. This makes it a perfect candidate for realistically modeling spatial extreme events over large and complex geographical domains. 

On one hand, desirable tail properties for our proposed model are achieved by carefully choosing the distribution of latent random factors and the structure of the latent processes. {A nice feature of our proposed model is that its marginal behavior and joint tail properties expressed through the $\chi$-coefficient can be expressed analytically in a simple and interpretable closed form. When $\beta=0$, the second-order tail-dependence coefficient $\bar\chi$ is either one or zero, depending on whether the two corresponding locations are covered or not by at least one compactly supported basis function used in the construction of the random scale process in our proposed SHOT model. We have provided detailed proofs of their derivations in the Supplementary Material. Computing the value of $\bar\chi$ when $\beta>0$ remains, however, an open question that we leave for future research.} On the other hand, computational speed is achieved by imposing a sparse probabilistic structure in the two components of our spatial-scale Gaussian scale mixture: we use the SPDE approach for the Gaussian process component, and a low-rank non-Gaussian process in terms of compactly supported basis functions for the random scale component. To further speed-up convergence and mixing of MCMC chains, our proposed Bayesian algorithm exploits customized adaptive block proposals for the latent variables. Moreover, non-extreme observations, which are treated as censored when they fall below a certain high marginal threshold, can be efficiently imputed thanks to the presence of the nugget effect, which allows us to reinterpret the data as being conditionally independent.

While the mesh of the SPDE approximation should be taken to be fine, there is a subtle trade-off between computational efficiency and approximate stationarity when choosing the number of basis functions for the random scale component. We suggest using a moderate number of basis functions and we have shown in our experiments how to strategically place the basis function knots (on the SPDE mesh) to mitigate undesirable nonstationary artifacts.

While spatial prediction is meaningless in our application (given that the domain is gridded and fully observed), it is also interesting to mention that the model allows for straightforward conditional simulation. This can be readily achieved at each iteration of our MCMC algorithm by noting that the process $Y(\bm s)=R(\bm s)W(\bm s)$ becomes Gaussian once we condition upon the random scale component $R(\bm s)$. This opens the door to extremal kriging with our new model. This is another advantage compared to the single scale-mixture HOT model, which returns spatial predictions similar to a Gaussian process as shown in \cite{Bolin.Wallin:2020}.

{In our real data application, we have used our model to study extreme monsoon precipitation data at 195 sites in Bangladesh, a relatively big country. Our results have shown that the SHOT model provides reasonably accurate fits for both margins and the extremal dependence structure across the domain, and outperforms natural competitors in terms of Bayesian model comparison through the deviance information criterion. While we compare the SHOT models with $\nu=1$ and $\nu=2$ in the Supplementary Material, considering $\nu$ to be an unknown parameter that can take non-integer values as in \cite{bolin2024covariance} and estimating it along with other parameters through a fully Bayesian estimation procedure, would be a future research endeavor. Furthermore, even though our study domain and dataset are already quite large and complex, it would be interesting in future research to test our model on even larger domains and bigger datasets, as its sparse structure allows it. However, the extremal dependence structure of our model remains approximately stationary. If the data require a nonstationary dependence structure, for example, if they are driven by mixtures of event types (e.g., convective vs frontal precipitation), a possible extension would be to consider a nonstationary Gaussian Markov random field model as in \cite{Lindgren.etal:2021} or Dirichlet process mixtures similar to \citet{hazra2018semiparametric}, \citet{hazra2019low} and \citet{Richards.etal:2023b}, but using the SHOT model for mixture components.}

\baselineskip=14pt
\bibliographystyle{apalike}

\bibliography{spatialextremes}
\end{document}